\documentclass[aps,pra,amsmath,amssymb,nofootinbib,superscriptaddress,notitlepage]{revtex4-1}

\usepackage[english]{babel}
\usepackage[utf8x]{inputenc}
\usepackage[T1]{fontenc}
\usepackage{siunitx}

\usepackage{graphicx}
\usepackage{booktabs}

\usepackage{xcolor}

\usepackage{mathtools}

\usepackage{physics}
\usepackage{rotating}
\usepackage{hyperref}

\begin{document}

\title{A case study against QSVT: assessment of quantum phase estimation improved by signal processing techniques}

\author{Sean Greenaway}
\email{Corresponding author: sgreenaway@psiquantum.com}

\author{William Pol}

\author{Sukin Sim}
\email{Corresponding author: ssim@psiquantum.com}

\affiliation{PsiQuantum, 700 Hansen Way, Palo Alto, California 94304, USA}

\begin{abstract}

In recent years, quantum algorithms have been proposed which use quantum phase estimation (QPE) coherently as a subroutine without measurement. In order to do this effectively,
the routine must be able to distinguish eigenstates with success probability close to unity. In this paper, we provide the first systematic comparison between two approaches towards maximizing this success probability, one using 
the quantum singular value transform and the other leveraging window functions, which have been previously studied as priors of the phase value distribution. 
We find that the quantum singular value transform is significantly outclassed by the window function approach, with the latter able to achieve 
between 3 and 5 orders of magnitude improvement in the success probability 
with approximately 1/4 the query cost. Our circuit simulation results indicate that QPE is not a domain which benefits from the integration of QSVT and we show that the use of the Kaiser window function is currently the most practical choice for realizing QPE with high success probability.

\end{abstract}

\maketitle

\section{Introduction}
Quantum computation holds the promise to accelerate calculations in a wide variety of domains for particular tasks, including search and optimization~\cite{grover1996fast}, simulation of chemistry~\cite{feynman1982simulating,georgescu2014quantum}, and cryptographic applications such as factoring prime numbers~\cite{shor1994algorithms, shor1997polynomial}. A ubiquitous workhorse algorithm used in all of these domains is the quantum phase estimation (QPE) algorithm~\cite{kitaev1995quantum, abrams1999quantum}, an algorithm which allows one to sample from the eigenspectrum of an input operator. Two common tasks which have exhibited computational speedups that make use of QPE include factoring prime numbers~\cite{shor1997polynomial, gidney2018factoring, litinski2023compute} and solving for the ground state energy of quantum mechanical systems in chemistry and condensed matter~\cite{whitfield2011simulation, reiher2017elucidating, babbush2018low, babbush2018encoding, Kivlichan2020improvedfault, lee2021even}.

For an operator $\hat{O}$ and one of its eigenstates $\ket{\psi}$ obeying the relation $\hat{O}\ket{\psi} = e^{2 \pi i \theta}\ket{\psi}$, QPE aims to output an estimate of $\theta$, the eigenphase corresponding to $\ket{\psi}$.\footnote{In our case, the operators of interest are typically Hamiltonians, so we will use the terminology ``eigenenergy'' and ``eigenvalue'' interchangeably.} Two salient criteria which characterize the performance and cost of QPE are the accuracy $\epsilon$ to which we estimate $\theta$, and the probability of success of returning an $\epsilon$-close estimate. Two factors influence this probability of success: the overlap of the initial state input to QPE with the true eigenstate $\ket{\psi}$, and more subtly, a phenomenon referred to as $\textit{bit discretization error}$, which arises from using a finite number $m$ bits to encode eigenphases and eigenenergies (which may have no exact finite-bit representation). These two factors contribute to the overall probability of success of returning an $\epsilon$-close estimate in different ways. The overlap of the initial state and $\ket{\psi}$ determines the likelihood of projecting onto $\ket{\psi}$ and returning a corresponding $m$-bit phase estimate. Assuming one successfully projects onto the eigenstate $\ket{\psi}$, bit discretization error determines the likelihood that a particular returned $m$-bit phase readout is $\epsilon$-close to the true target eigenphase $\theta$. Addressing the influence on probability of success due to bit discretization error is an ongoing line of research in the literature and the focus of this study.

Two methods have emerged in the literature that aim to address bit discretization error: the quantum singular value transform (QSVT) framework applied to QPE \cite{gilyen2019quantum, rall2021faster,martyn2021grand}, which we will refer to as ``QSVT QPE,'' and the use of window or taper functions \cite{babbush2018encoding, rendon2022effects, patel2024optimal}. The QSVT framework is a powerful conceptual tool, as it recasts many tasks in quantum computation as attempting to implement a function on the eigenvalues or singular values of an operator (using the language of signal processing). Window functions are tools borrowed from classical signal processing
that are used to reduce errors in the frequency spectra of signals. 
Here we compare and contrast the relative success probabilities and resource costs of these two methods.

The favorability of one method over the other depends on how QPE is used in a full computation, and the requirement on the success probability. It is useful to distinguish between $\textit{coherent}$ and $\textit{incoherent}$ uses of QPE. In an incoherent use of QPE, we measure the phase qubits register immediately after performing a single iteration of QPE. In contrast, when we use QPE coherently, we do not measure the phase qubits register in order to avoid collapsing the resulting superposition state. For some applications, maintaining coherence is necessary \cite{hhl, metropolis_sampling, yung2012quantum, Lemieux_2020, Montanaro_2015, harrow2020adaptive, Arunachalam_2022}. One such application requiring high probability and coherent use of QPE is estimating the expectation value of an observable $\hat{F}$ with respect to a state $\ket{\psi}$ that is $\textit{not}$ its eigenstate \cite{steudtner2023fault}. This application requires repeated application of a unitary that performs a reflection about the state $\ket{\psi}$. Constructing such a reflection is in general non-trivial; instead, QPE is used to construct a reflection about $\ket{\psi}$'s corresponding eigenphase as a proxy. Errors due to inexact, approximate implementations of reflections play a significant role in the overall success probability of algorithms that make repeated queries to reflections. With repeated queries, error will accrue during the course of the computation, as in amplitude amplification or amplitude estimation. For the reflection implemented in \cite{steudtner2023fault}, this QPE-based reflection only approximately performs (a proxy for) a reflection about $\ket{\psi}$, precisely because the exact eigenphase is not representable with a finite number of bits. Mitigating bit discretization error and its influence on the success probability of QPE is crucial to this kind of application.

Many of the conclusions we draw on the relative advantage in performance between the use of window functions in QPE and QSVT QPE in the rest of this manuscript are particular to regimes in which the required success probability is close to $1$. For the rest of this manuscript, we explore the tradeoffs of window functions and QSVT QPE assuming such a regime. One may consider the coherent use of QPE for calculating observables of operators with respect to bases that do not form part of the operator's eigenbasis (as mentioned above) as a prototypical application. Ultimately, it is found that for this high-accuracy, high-success probability regime, it is more advantageous and less costly to use window functions than it is to use QSVT methods. 
In Section~\ref{sec:qpe_developments}, we detail recent developments for both methods and their use in QPE. In Section~\ref{sec:sim_results}, we detail the numerical circuit simulation results comparing the impact of these options on the total success probability and resource cost of the algorithm. Finally, we comment on the broad utility of methods like QSVT for QPE in the conclusion.

\section{Recent developments on QPE}
\label{sec:qpe_developments}

Performance metrics for QPE have been extensively studied in the literature \cite{luis1996optimum, dam2007optimal, babbush2018encoding, najafi2023optimum}. Choosing an appropriate cost function (e.g. minimizing the mean-square-error or the Holevo variance) is highly dependent on the task-at-hand, i.e. estimating the eigenphase from measurements or using the QPE coherently as a subroutine. 
Among these metrics, the \emph{success probability} of QPE is
an important cost function for cases in which a higher-level algorithm or application uses QPE coherently. 
In our work, we define the success probability of QPE as the sum of probabilities corresponding to the two nearest $m$-bit 
fixed-point approximations to the true eigenphase, corresponding to the probability of obtaining an $m-$bit estimate $\varphi$ of the true eigenphase $\phi$ of a unitary $U$ such that $|\varphi - \phi| \leq \frac{1}{2^m}$.\footnote{The two closest binary strings are equally good approximations only in the case that the true phase lies precisely halfway between them: for many phases there will be an unambiguous closest $m-$bit approximation to the true phase $\phi$. This satisfies a slightly different inequality $|\varphi - \phi| \leq \frac{1}{2^{m+1}}$. Typically the number of bits $m$ is chosen such that the precision set by $\frac{1}{2^m}$ is sufficient and hence measuring either the closest or second closest bitstring is acceptable.} For the textbook implementation of QPE, this success probability is bounded by $P_{succ} \geq \frac{8}{\pi^2} \approx 0.81$~\cite{cleve1998quantum,nielsen_chuang_2010}.
One way to verify this notion of success probability is by simulating the output probabilities of the $m$ phase qubits used in QPE. This results in a histogram of $2^m$ bins, in which each bin in the histogram corresponds a bitstring representation encoding a binary fixed-point number. In practice, we take a finite number of measurements, resulting in an estimated histogram.

A standard method of boosting the success probability given an instance of QPE is to add to the number of phase qubits $m$ then ignore the additional qubits.\footnote{This method can be applied in the case of the reflection about the ground state mentioned above by making use of a multi-controlled Z gate with zero and one controls corresponding to the binary representation of the previous $m$-bit estimation of the eigenphase. For more detail, readers should refer to Ref.~\cite{steudtner2023fault}.}
That is, adding phase qubits fine-grains the histogram of QPE outcomes and doubles the number of bins for each additional qubit. By ignoring the additional qubits, say $p$ of them, this is effectively coalescing the bins from $2^{m+p}$ to the original $2^{m}$ bins and summing up the probabilities from the combined bins.
In this section, we describe two alternatives, more recent developments that boost the success probability of QPE, namely the use of window or taper functions \cite{babbush2018encoding, rendon2022effects, berry2022quantifying} and the quantum singular value transform (QSVT) \cite{rall2021faster, martyn2021grand}.

\subsection{Quantum phase estimation using window functions}

\begin{figure}[h]
\centering
\includegraphics[width=0.8\textwidth]{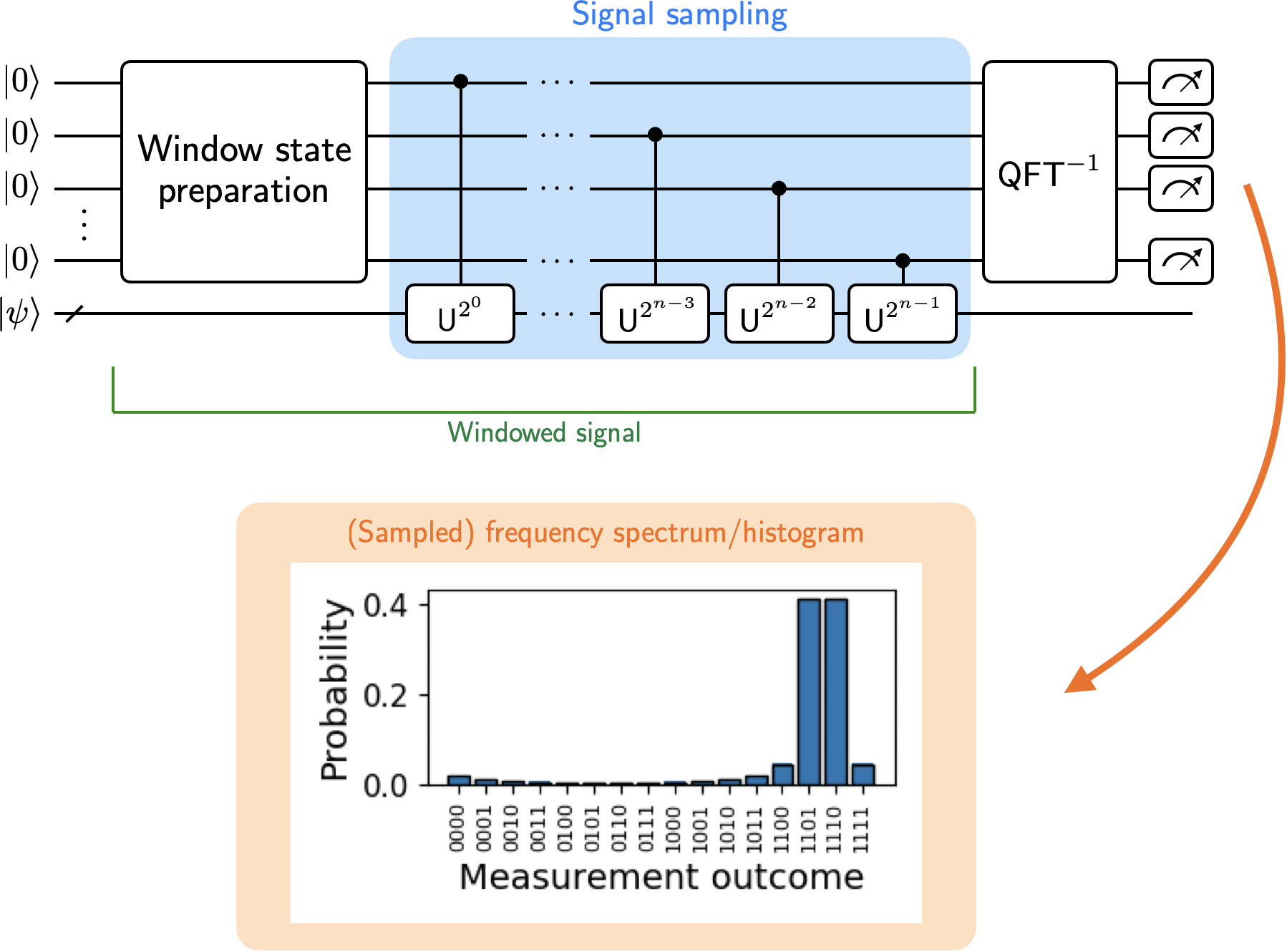}
\caption{Quantum phase estimation re-interpreted using signal processing language. A window state is first prepared on the phase register. Next, the signal is sampled (in time) by applying controlled-unitaries that contain information about the true phase, resulting in preparation of a windowed signal. Following the inverse quantum Fourier transform and measurement, we obtain a \emph{histogram} of the QPE outcomes corresponding to the frequency spectrum of the windowed signal by repeated measurement of the phase register.
By choosing an appropriate window function, the output spectrum can have reduced spreading of frequencies which appear due to time-limited sampling and windowing.}
\label{fig:windowed_qpe_signal_processing}
\end{figure}

Studies of window states in QPE originate from ideas in quantum metrology \cite{luis1996optimum, dam2007optimal, babbush2018encoding, najafi2023optimum}, where they were often referred to as \emph{control states}. 
The optimal control state for minimizing a particular cost function of QPE, namely the Holevo variance, was derived by Luis et al \cite{luis1996optimum}.\footnote{In this work, we refer to this control state as the ``sine window'' as shown in Table~\ref{tab:window_functions}.}
In this section, we summarize an alternative yet likely related interpretation of QPE in the language of signal processing, as illustrated in Fig.~\ref{fig:windowed_qpe_signal_processing} and as is suggested, for instance, in Section II of \cite{rendon2022effects}. We first motivate window functions in signal processing before describing their connection to QPE.

In signal processing, we may observe \emph{spectral leakage} or introduction of new frequency components when we take the Fourier transform of a time-limited (or truncated) signal \cite{harris1978use}. 
By truncating in the time domain, one may introduce discontinuities at the ends of the signal. The Fourier transform assumes the sample to be periodic and will thus introduce unexpected frequency components due to the discontinuities.
To reduce this leakage, a \emph{window} or \emph{tapering} function is chosen and multiplied to the sampled signal. 
A window function is often tapered at its ends, thus multiplying a signal by this function smooths the discontinuities at the ends of the sampled signal.
However, a window function is characterized by its own set of frequencies and therefore, multiplying the window function and the sampled signal in the time domain corresponds to convolving in the frequency domain, resulting in some spreading of frequencies in the output spectrum.

In the textbook version of quantum phase estimation, as illustrated in Fig.~\ref{fig:windowed_qpe_signal_processing}, we first apply Hadamard gates on the phase register of size $m$, preparing a uniform superposition.
This, in fact, realizes a rectangular or ``boxcar'' window function on the amplitudes.
However, we can replace the Hadamards with a different state preparation unitary.
We will call the resulting state a \emph{window state} but will use this term and ``window function'' interchangeably.
After preparing the window state and assuming the input eigenstate $\ket{\psi}$ has already been (perfectly) prepared, a sequence of controlled-$U^{2^k}$ is applied where $U\ket{\psi} = e^{2\pi i \phi}\ket{\psi}$ and $k$ runs from 0 to $m-1$.\footnote{Note that the window state in QPE has an additional role of preparing a state that can activate the controls in the controlled-$U^{2^k}$ gates. Thus, the window state preparation must come before the application of controlled-unitaries that corresponds to preparing a (time-limited) signal. 
Alternatively, one could view sending $\ket{0}$ into QPE as preparing a poor choice of window function.}
Lastly, the inverse quantum Fourier transform (QFT) is applied to the signal to obtain 
the histogram of measurement outcomes.
In practice, this histogram is estimated by taking some finite number of measurements. 

In the case that the true phase $\phi$ is expressible in $m$ bits or fewer, the signal is periodic on the interval, where we have assumed a rectangular window state. Thus the phase value can be extracted after applying the inverse QFT.
However, for generality, we assume $\phi$ to be an irrational number.
In such cases, applying up to controlled-$U^{2^m}$ leads to preparing a time-limited signal that is not periodic on the interval. Applying the inverse QFT on the time-limited sampled signal results in a frequency spectrum that has a spread about the true phase value due to spectral leakage.
To reduce this leakage, as is done in signal processing, we can consider preparing an alternative window state on the phase register.
In the literature, several window states have been proposed: cosine \cite{rendon2022effects}, sine \cite{luis1996optimum,dam2007optimal,babbush2018encoding}, Kaiser \cite{berry2022quantifying}, and DPSS \cite{patel2024optimal} windows, also summarized in Table~\ref{tab:window_functions}. It was shown that of the aforementioned windows, the Kaiser window has the most favorable scaling in terms of the number of additional phase qubits required in QPE to achieve an error to within $1/2^{m}$ \cite{berry2022quantifying}.
We provide a sketch of deriving the number of additional qubits for the Kaiser window in Appendix~\ref{app:additional_qubits_kaiser}.

\begin{figure}[h]
\centering
\includegraphics[width=0.8\textwidth]{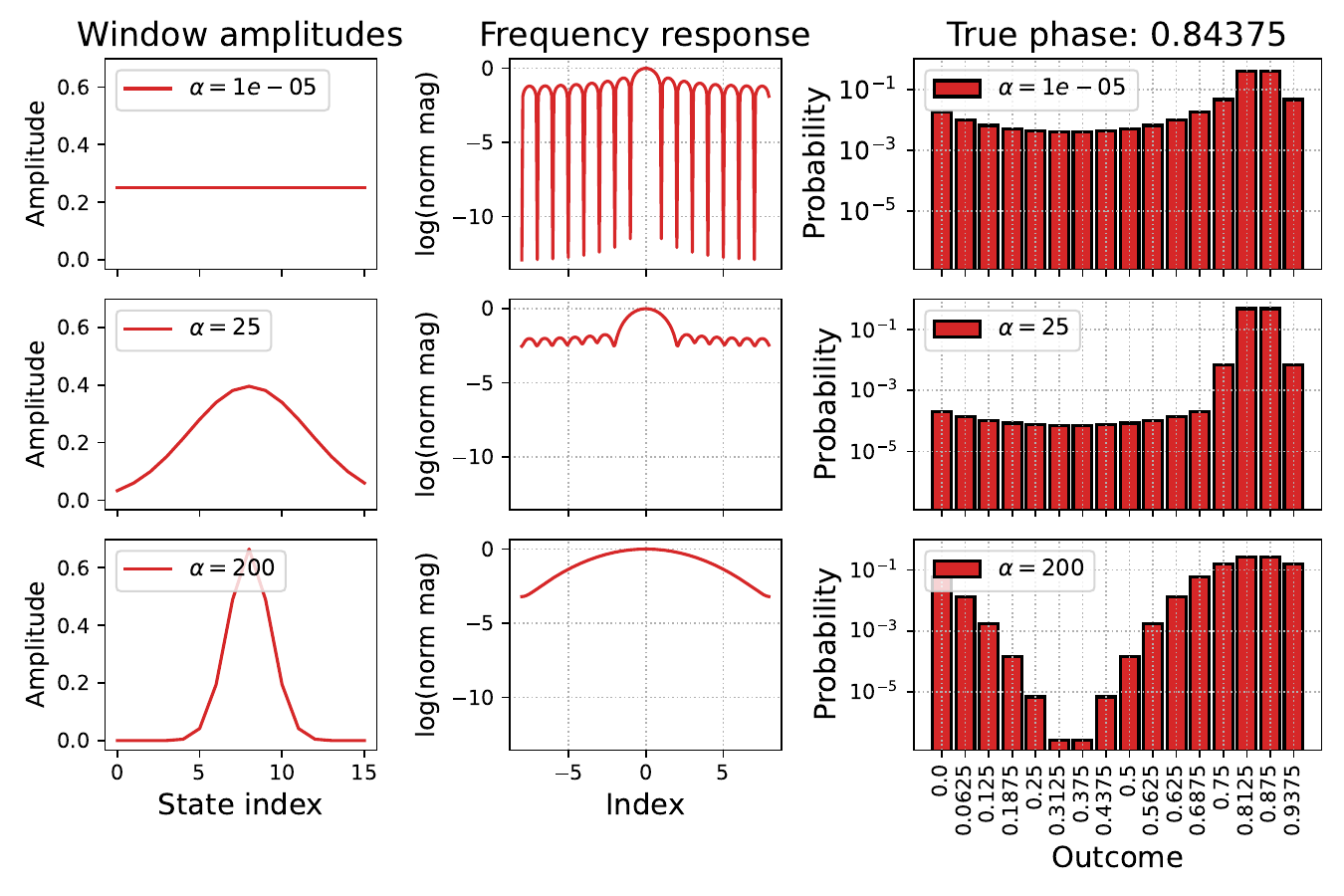}
\caption{Numerical analysis of the Kaiser window for QPE.
This toy QPE example uses four bits of precision, and the true phase is 0.84375, which requires 5 bits to express. This value lies exactly in between two bins (or bitstrings) representing 0.8125 and 0.875.
We investigate three different instances of the Kaiser window, in which the parameter $\alpha$ is varied. 
The first column shows the window function amplitudes for a four-qubit state (i.e. the QPE uses four bits of precision).
The second column shows the frequency spectrum of each window state, in which one can see that varying $\alpha$ affects the side lobes in the Fourier spectrum. 
The third column shows the output probabilities of phase values from simulating the QPE circuit.}
\label{fig:kaiser_window_analysis}
\end{figure}

To illustrate
the capabilities of the Kaiser window, we consider a toy example of a four-qubit QPE in Fig.~\ref{fig:kaiser_window_analysis} with the true phase value that sits equally between two bins.
The Kaiser window state is defined as follows:
\begin{align}
    \ket{\psi}_{\textsf{Kaiser}} = \sum_{x=-2^{m-1}}^{2^{m-1}} \frac{1}{2^m} \frac{I_0(\pi \alpha \sqrt{1-(x/2^{m-1})^2})}{I_0(\pi \alpha)} \ket{x},
\end{align}
\noindent where $I_j(\cdot)$ is the modified Bessel function of the first kind of order $j$.
This window state has a parameter $\alpha$ that can 
be used to 
tune the tapering capability, as shown in the first column of Fig.~\ref{fig:kaiser_window_analysis}.
When $\alpha$ is very small, as in the first row of the figure, this approaches the rectangular or boxcar window state.
When $\alpha$ is large, e.g. 200, this corresponds to a fast-decaying taper function. This raises the question of how to select the right $\alpha$ for the Kaiser window or more broadly how to choose an appropriate window state. Fortunately, we can refer to existing techniques in signal processing for analyzing and choosing an appropriate window state.

In the field of signal processing, 
several families of window functions have been proposed.
These functions can be evaluated by analyzing their frequency spectra, extracting properties such as width of the main lobe or the height and decay rate of the side lobes. In the middle column of Fig.~\ref{fig:kaiser_window_analysis}, we show the frequency spectra of Kaiser windows with different $\alpha$ values. The main lobe corresponds to the 
central peak. 
Main and side lobes of a window function can help inform the degree of spreading of frequencies expected in the final frequency spectrum, which in our context is the histogram of QPE outcomes. 
For instance, using a window state with a wide main lobe results in a relatively wide 
spread 
of frequencies (or phase outcomes) about the true frequency (phase). On the other hand, a tall and slowly decaying side lobe results in spreading of frequencies throughout, including frequencies further away from the true phase.
In practice, there is a trade-off between main lobe width and side lobe height.
In the last column of Fig.~\ref{fig:kaiser_window_analysis}, we show the outcomes of the four-qubit QPE produced using state vector simulations.
We can compute the success probability of this QPE by summing up the heights or probabilities of bins corresponding to 0.8125 and 0.8750.
The frequency spectrum of the Kaiser window where $\alpha=10^{-5}$ (effectively a rectangular window) 
features a relatively narrow main lobe but tall and slowly decaying side lobes. Thus, in the corresponding QPE histogram, this results in some spreading of phase values away from the true phase.
On the other end, when $\alpha$ is large, e.g. 200, the main lobe dominates, and we observe significant concentration (though broadened) of phase values about the true phase.
If we set $\alpha$ to $25$, the main lobe width is 
similar to the case where $\alpha=10^{-5}$, but the side lobe heights are decreased. Applying this window state reduces the spreading of phase values further away from the true phase.
We note that of the three cases, the QPE using the Kaiser window with parameter value $\alpha=25$ has the highest success probability \emph{as well as} reduced side lobes.
This demonstrates the importance of tuning $\alpha$ to an appropriate value for the window state to 
be 
effective. 
We re-visit the discussion on choosing an appropriate $\alpha$ in Appendix~\ref{app:sim_details_windows}.

Lastly, in the context of QPE, we must also consider the cost of preparing window states.
Fortunately, costs of the sine and cosine window functions are considerably low, as discussed in \cite{babbush2018encoding} for the sine window. We show the circuit for preparing the cosine window state in Fig.~\ref{fig:cosine_qpe_circuit}, in which we expect the total cost of the controlled-unitaries $U$ or block-encodings to be significantly higher.
We also provide some rough cost estimate for the Kaiser window based on previous works in Appendix~\ref{sec:envelope} and again show that its cost may be insignificant compared to the total costs of the controlled-unitaries or block-encodings.

While window functions in QPE are well-motivated, and several previous studies have focused on their asymptotic performance \cite{patel2024optimal}, there is still a lack of understanding in how they compare against other existing improvements to QPE such as the use of the quantum singular value transform (QSVT). In the following section, we briefly outline the theory behind QSVT-QPE before comparing the numerical performance of the two approaches via state vector simulations.

\begingroup

\begin{table}[]
\setlength{\tabcolsep}{8pt} 
\renewcommand{\arraystretch}{2.5} 
\centering
\begin{tabular}{|c|c|c|}
\hline
\textbf{Name/Ref} & \textbf{Window function form} & \textbf{Additional qubits} \\ \hline
Rectangular & $\sum_{x=0}^{2^{m}-1}\frac{1}{\sqrt{2^m}} \ket{x}$ & $\mathcal{O}(\log(1/\delta))$ \cite{cleve1998quantum} \\ \hline
Cosine \cite{rendon2022effects} & $\sum_{x=-2^{m-1}}^{2^{m-1}} \frac{\sqrt{2}\cos(\frac{\pi x}{2^m})}{\sqrt{2^m}} \ket{x}$ & $\mathcal{O}\Big(\log(1/\delta^{1/3})\Big)$ \\ \hline
Sine \cite{babbush2018encoding} & $\sum_{x=0}^{2^{m}-1} \sin \big( \frac{\pi(x+1)}{2^m + 1} \big) \ket{x}$ & Similar to cosine window \\ \hline
Kaiser \cite{berry2022quantifying} & $\sum_{x=-2^{m-1}}^{2^{m-1}} \frac{1}{2^m} \frac{I_0(\pi \alpha \sqrt{1-(x/2^{m-1})^2})}{I_0(\pi \alpha)} \ket{x}$ & $\mathcal{O}(\log \log(1/\delta))$ \\ \hline
\end{tabular}
\caption{Summary table of select window functions. Here, $m$ is the number of phase qubits.
Depending on the window function expression, the index $x$ runs from either $0$ to $2^{m}-1$ or $-2^{m-1}$ to $2^{m-1}$, but their corresponding basis states are the same.
The Kaiser window function has a tunable parameter $\alpha$ that can balance the effects of the main lobe width and side lobe height in the frequency spectrum. The function forms of these window functions may be un-normalized. The column ``Additional qubits'' shows the number of phase qubits one could add then discard to achieve 
a success probability of 
$1 - \delta$, 
i.e. the probability of getting an error within $\frac{1}{2^{m}}$. 
We expect the additional qubits required for the sine window to be similar to that of the cosine window. We show a sketch of the proof for the number of additional qubits for the Kaiser window in Appendix~\ref{app:additional_qubits_kaiser}.
}
\label{tab:window_functions}
\end{table}

\endgroup

\subsection{Quantum phase estimation using QSVT}
As shown in the previous section, the success probability of QPE with respect to bit discretization error can be boosted by making an appropriate choice of prior distribution over the states of the phase qubits before performing the phase kickback. This strategy does not alter the eigenphase kicked back by the controlled unitaries in QPE, but rather uses the statistical properties of the state stored in the phase register to achieve the higher success probability. An alternative approach would be to instead alter the unitary whose eigenphase is being kicked back such that the bit discretization error is reduced or (ideally) eliminated. Intuitively, we wish to implement a modified unitary $\tilde{U}$ whose eigenvalues correspond to those of the ideal unitary, truncated precisely to $m$ bits of precision such that the bit discretization error of the modified unitary is zero.

Such a modified unitary can be realized as a matrix function of the original target unitary $U$, which itself may be implemented using the {\it quantum singular value transform} (QSVT), a protocol for implementing matrix polynomials on a quantum computer -- for the interested reader, a brief overview of QSVT is provided in Appendix~\ref{sec:qsvt_overview}. The main idea behind QSVT is to implement a block encoding of some input unitary
\begin{equation}
    \begin{bmatrix}
    f(U) & * \\
    * & *
    \end{bmatrix} \ ,
\end{equation}
where the elements labelled $*$ are left undefined and where the function $f$ acts on the singular values of $U$. In Ref.~\cite{martyn2021grand}, the authors present a method for using QSVT to boost the success probability of QPE, a routine that we refer to hereafter as {\it QSVT QPE}. The core of that routine revolves around using QSVT to implement the shifted sign function
\begin{equation}
    f(x) = \Theta\left(\frac{1}{\sqrt{2}} - x\right) \ ,
\end{equation}
with $\Theta(x)$ denoting the sign function, using the notation of Ref.~\cite{martyn2021grand} (not to be confused with the Heaviside step function). At a high level, one can think of this function as performing the transformation described above, clipping phases that cannot be represented by $m$ bits such that the transformed eigenphase can be represented exactly using $m$ bits. For the interested reader, a detailed overview of the choice of function is given in Appendix~\ref{app:qsvt_function_choice}.

One subtlety is that this construction requires that at each iteration, the less significant bits in the eigenphase i.e. those that were previously measured in the QPE protocol, should be rotated out from the unitary so they are not measured again. In textbook QPE, this is implemented by the controlled phase rotations in the inverse quantum Fourier transform. This is a well-known technique that underpins {\it iterative} quantum phase estimation~\cite{cleve1998quantum}, although unlike iterative methods, the routines we consider here are performed coherently. The specific circuit details for implementing this algorithm are outlined in Appendix~\ref{app:sim_details}.

\subsection{Challenges of QSVT in practice}
The principle challenge associated with implementing routines based on QSVT lies in obtaining the phase factors that implement a given function. There is no closed-form expression for these phase factors for a given general target function, and so numerical methods must be used to obtain them. There are two approaches for this: optimization-based~\cite{dong2021efficient} and non-optimization-based methods ~\cite{gilyen2019quantum,haah2019product,chao2020finding,ying2022stable}. In practice, we find that the non-optimization-based methods are less numerically stable and do not provide a significant advantage in terms of accuracy or speed compared with optimization-based methods, so we use optimization in this work. For the numerical results presented here, we used a simple \texttt{scipy} minimization of the absolute mean squared error between the signal processed unitary matrix and the target function; we also evaluated the performance of the QSPPACK library~\cite{qsppack}, but found that it did not make any significant difference to the results, possibly due to the relatively low degrees (and therefore relatively simple optimization landscape) explored herein.

The typical workflow for implementing a function in QSVT is to use the function we wish to implement as the target for an optimization, with the phase angles as the optimization parameters. However, since the shifted sign function is discontinuous, an additional decomposition step must be made to obtain a continuous and symmetric approximation to the sign function. The decomposition used in this work~\cite{martyn2021grand}\footnote{Note that our notation differs slightly from that presented in Ref.~\cite{martyn2021grand} – our $\kappa$ corresponds to their $\Delta$, and our $\Delta$ corresponds to their $\epsilon$.} is
\begin{equation}\label{eq:poly_approx}
    P_{\Delta, \kappa}(x) := -\frac{1}{1 + \frac{\Delta}{4}}\left(-1 +\frac{\Delta}{4} + P^\Theta_{\frac{\Delta}{2}, \kappa}\left(\frac{1}{\sqrt{2}} - x\right) + P^\Theta_{\frac{\Delta}{2}, \kappa}\left(\frac{1}{\sqrt{2}} + x\right)\right) \ ,
\end{equation}
where
\begin{equation}
    P^\Theta_{\frac{\Delta}{2}, \kappa}(x) := \erf\left(\frac{\sqrt{2}}{\kappa}\sqrt{\log\left(\frac{2}{\pi\Delta^2}\right)} x\right) \ .
\end{equation}
These functions have two parameters which control the quality of the approximation: $\Delta$, which controls the maximum deviation away from the true function and $\kappa$, the region around the discontinuity where the error is allowed to be arbitrarily large. These two parameters control the quality of the approximation, and typically the lower their values, the higher the polynomial degree needed to implement it (and hence the higher the gate cost). In Ref.~\cite{martyn2021grand}, it is shown that a value of $\kappa < 2\left(\cos\left(\frac{3\pi}{16}\right)-\frac{1}{\sqrt{2}}\right) \approx 0.25$ suffices to yield a high success probability. For $\Delta$, it is shown that a value of $\Delta \leq \sqrt{2\delta/(m+1)}$ (where $m$ is the number of bits of precision) is required to obtain a success probability greater than $1 - \delta$. The choice of the polynomial degree for the QSVT function acts as a parameter establishing a tradeoff between cost and performance, analogous to the number of additional bits $p$ used in window functions.

Obtaining values for the function parameters is not the end of the story, however. One must then obtain the optimal phase factors that realize this function. In practice, while we find that the analytical values provide a good starting point for an optimization, it is often useful to vary these values to obtain as small a value for $\Delta$ as possible for a given degree. In this work, the QSVT QPE results correspond to fully compiled decompositions with the phase angles obtained through such an optimization.

\section{Simulation results}
\label{sec:sim_results}
In order to evaluate and compare the impact of window functions and QSVT on QPE, it is useful to consider numerical state vector simulations of the full subroutine, which we implement 
using a proprietary state vector simulator.
The details of the specific circuits used in this work (including corrections to the original QSVT QPE circuit) are outlined in Appendix~\ref{app:sim_details}. Throughout, we use a simple $\text{phase}$ gate to generate the eigenphase we wish to read:
\begin{align}\label{eq:Pgate}
    P(\phi) := \begin{bmatrix}
1 & 0\\
0 & e^{2\pi i\phi}
\end{bmatrix}.
\end{align}
The simplicity of this unitary does not affect the generality of the results presented here -- the reduction in success probability from unity arises due to bit discretization errors associated with representing the eigenphase with a finite number of bits (assuming, as we do throughout, that we are working with an exactly-implemented input eigenstate). As a result, the same phenomenon will occur regardless of which unitary gives rise to the eigenphase being measured, and therefore the results we present here should be applicable even for large-scale applications. The specific system being investigated will impact the precision required to obtain a result to a desired accuracy (for instance, unitary encodings of Hamiltonians with large norms require more bits of precision than those with smaller norms to achieve the same accuracy), but this is a separate concern from the core question investigated here -- explicitly, the setting for this work is: given sufficient bits of precision $m$ to achieve a desired accuracy $\epsilon < 1/2^{m}$, how many {\it additional} resources are needed to achieve a high probability that the measured phase is $\epsilon-$close to the true phase?

There are two additional considerations that need to be pinned down in order to obtain concrete numerics: the desired success probability and the method by which it is evaluated over the full range of possible eigenphases. Since the performance of any QPE routine will vary (often significantly) for different target eigenphases, this analysis only makes sense if we consider the full range of phases when deciding if the success probability is sufficiently high. In real applications, we will not know the true eigenphase (this is one of the major reasons for doing QPE in the first place!) and so we cannot know if our target lies in some fortunate, highly performant region of the possible phases, or if it happens to be some pathological case with significantly lower success probability. As such, we consider the minimum success probability over this full range to be the most important value for these numerics. Here, we choose the additional resources (either additional phase qubits for the window function QPE routines or the polynomial degree in the case of QSVT QPE) to be the minimal resources necessary to achieve a minimum success probability of 99\%. This value is, in some sense, arbitrary, but we find our results remain consistent for other values.

\subsection{Success probability}

\begin{figure}[h]
\centering
\includegraphics[width=0.6\textwidth]{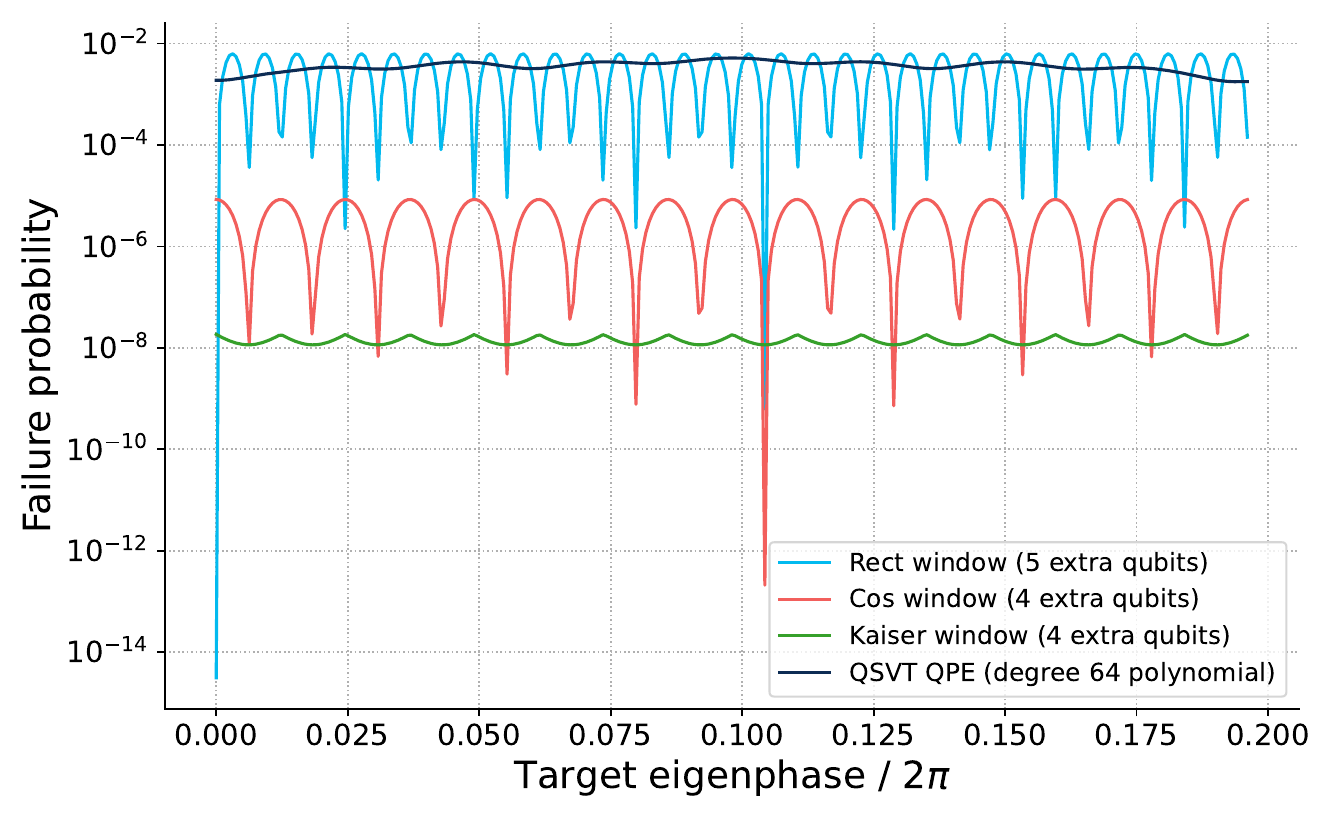}
\caption{Success probability as a function of the measured eigenphases for quantum phase estimation with different modifications. The dark blue line represents QSVT QPE with a $d=64$ degree polynomial as outlined in Ref.~\cite{martyn2021grand}, and the sky blue, red, and green lines represent QPE with rectangular, cosine and Kaiser windows respectively. For the rectangular window, $p=5$ additional qubits were used, while for the cosine and Kaiser windows $p=4$ additional qubits were used; for QSVT QPE no additional phase qubits were used. All the QPE routines use $m=5$ phase qubits to load the approximation of the eigenphases.
}
\label{fig:success_probs_5_bits}
\end{figure}

In this section, we present results from numerical experiments, highlighting the relative success probabilities of using window states or QSVT in QPE.
Fig.~\ref{fig:success_probs_5_bits} shows the success probabilities for different five phase bit QPE implementations as a function of the $P(\phi)$ eigenphase.
As previously motivated, the primary figure of interest for this work is the minimal success probability (or equivalently, one minus the maximum failure probability) achieved over the full range of possible eigenphases. By this metric, the best implementation by a wide margin is the QPE using a Kaiser window function and four additional phase qubits, achieving a maximum failure probability of $10^{-7.28}$, followed by QPE using a cosine window and four additional qubits, achieving a maximum failure probability of $10^{-5.07}$. QSVT QPE achieves a maximum failure probability of $10^{-2.28}$, with the rectangular window function having the worst performance, achieving a maximum failure probability of $10^{-2.2}$.

\subsection{Costs of the different implementations}

From the results in Fig.~\ref{fig:success_probs_5_bits}, one may conclude that all considered QPE implementations achieve very high success probabilities. The Kaiser window appears to achieve the highest success probabilities over possible values of the target eigenphase, but other implementations also achieve values close to 1.
However, to establish a fair comparison, we additionally provide the relative costs of different QPE implementations, specifically the numbers of unitaries called in QPE.
We assume that the unitary is a {\it block encoding} $U_A$ of some matrix $A$, defined as
\begin{equation}
    U_A := \begin{bmatrix}
A & *\\
* & *
\end{bmatrix} \ ,
\end{equation}
where the elements labelled $*$ are left undefined. 
This block encoding is 
representative of the unitaries that are commonly used in fault-tolerant quantum algorithms that make use of QPE \cite{babbush2018encoding, lee2021even}.

\begin{figure}[h]
    \centering
    \includegraphics[width=0.6\textwidth]{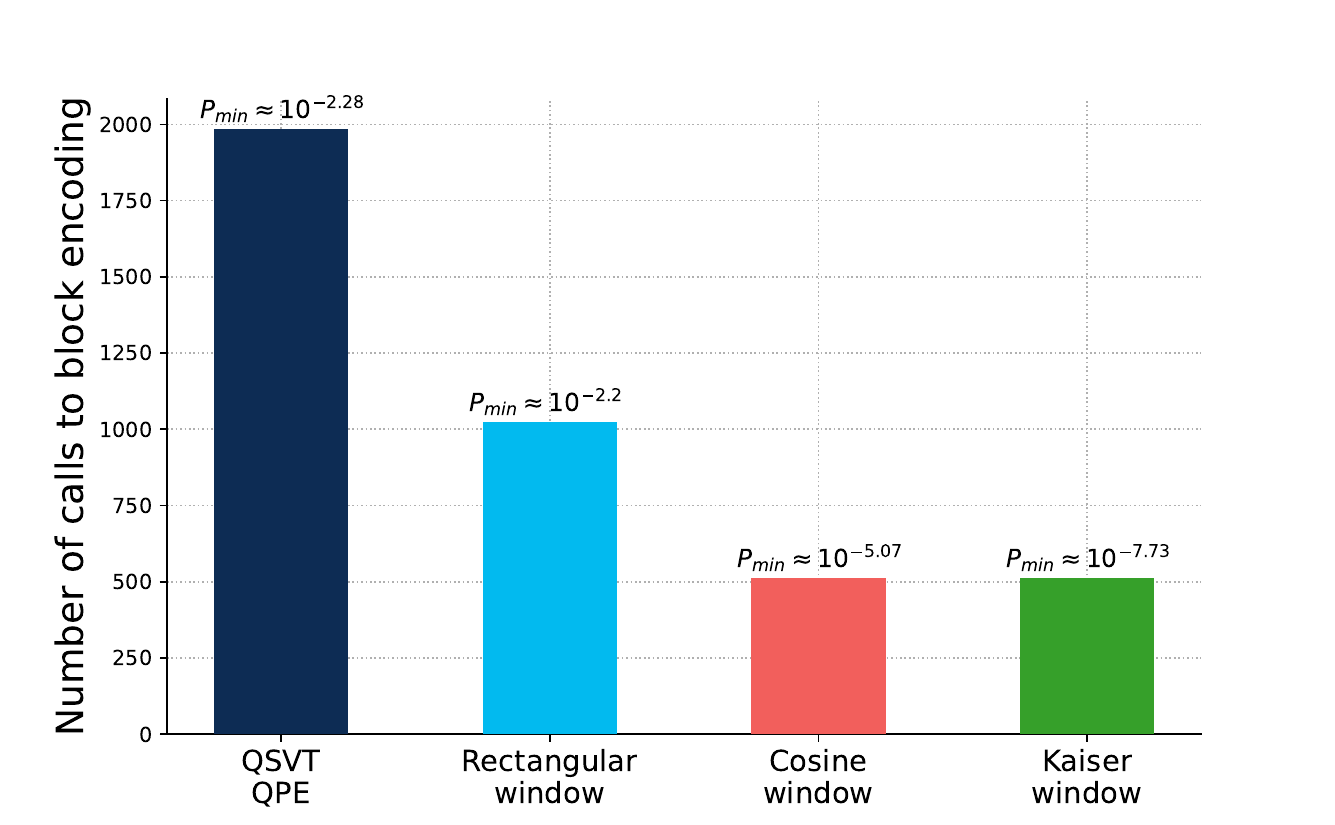}
    \caption{Graph of the cost in terms of queries to the block encoding unitary for QSVT QPE with a degree $d=64$ polynomial (green), vanilla QPE with a rectangular window function (dark blue), vanilla QPE with a cosine window function (red) and vanilla QPE with a Kaiser window function (sky blue). All results are for $m=5$ bit quantum phase estimation routines, with the degree of the polynomial approximation (for QSVT QPE) and the number of additional phase qubits (for the window functions) chosen such that the minimal success probability was greater than 0.99. Numbers above the bars are the maximum failure probability, or one minus the minimum success probability, over possible values of the target eigenphase (see Fig.~\ref{fig:success_probs_5_bits}).}
    \label{fig:be_cost_5_bits}
    \end{figure}

Fig.~\ref{fig:be_cost_5_bits} gives some proper context to the results shown in Fig.~\ref{fig:success_probs_5_bits}: while QSVT QPE performs fairly well in terms of minimal success probability, its cost is by far the largest out of the implementations considered, at almost 1984 calls to $U_A$. The rectangular window function is also costly, albeit much less so than QSVT QPE, coming in at 1023 calls to $U_A$. By far the cheapest routines were the cosine and Kaiser window function implementations, with 127 calls to $U_A$ each. These low costs reflect the fact that only 4 additional phase qubits are needed to achieve a success probability $\geq 0.99$, as opposed to the 5 that are required for the rectangular window implementation. It should be noted that, asymptotically, the Kaiser window should perform exponentially better than the cosine window, scaling as $\log\log (1/\delta)$ as opposed to $\log (1/\delta)$ as shown in Appendix~\ref{app:additional_qubits_kaiser}. However, since the number of phase qubits required by the cosine window implementation is relatively low, at only 4 additional phase qubits, this asymptotic advantage is not
yet
realized 
(i.e. the Kaiser window with only 1 additional phase qubit failed to achieve a consistent success probability greater than $0.99$), 
likely 
due to the neglected constant factors in the asymptotic expressions. 
We show the crossover point between the two window functions in the following section.

For QSVT QPE, the cost is significantly higher due to the number of calls to $U_A$ required to implement the sign function. This disparity of costs is unlikely to be overcome simply by obtaining a better polynomial approximation or performing a better optimization, which can be seen by explicitly calculating the maximum allowable degree as constrained by the cosine and Kaiser window costs.
The cost for both these implementations is $2^{m+p}-1=2^{5+4}-1=511$. In order to yield a QSVT QPE routine with a total cost less than this, we can set the degree to $d = \lfloor 127/(2^m-1)\rfloor=\lfloor 511/31\rfloor=16$. Thus, in order for QSVT QPE to be competitive with the cosine and Kaiser window functions, a degree-16 approximation to the sign function is required to yield a maximum failure probability less than $10^{-7}$. Even with improvements to the QSVT QPE protocol, it seems unlikely that this performance gap can be overcome given the additional resources available.

\subsection{Pushing the success probability even higher}
The above analysis shows that the use of window functions (and in particular the Kaiser window function) yields QPE routines with higher success probability than QSVT QPE, but the chosen value of 0.99 for the success probability may not be indicative of the performance of these subroutines in real applications. For algorithms that make use of coherently controlled QPE as a subroutine, there is often a correlation between the success probability of the QPE routine and the performance of the overall algorithm (although this is not the only factor to consider when designing such an algorithm). A pertinent question is therefore whether the window functions maintain their higher performance over QSVT QPE at higher success probabilities, or whether there is some cross-over point after which it becomes favorable to utilize QSVT QPE.

\begin{figure}[h]
\centering
\includegraphics[width=0.6\textwidth]{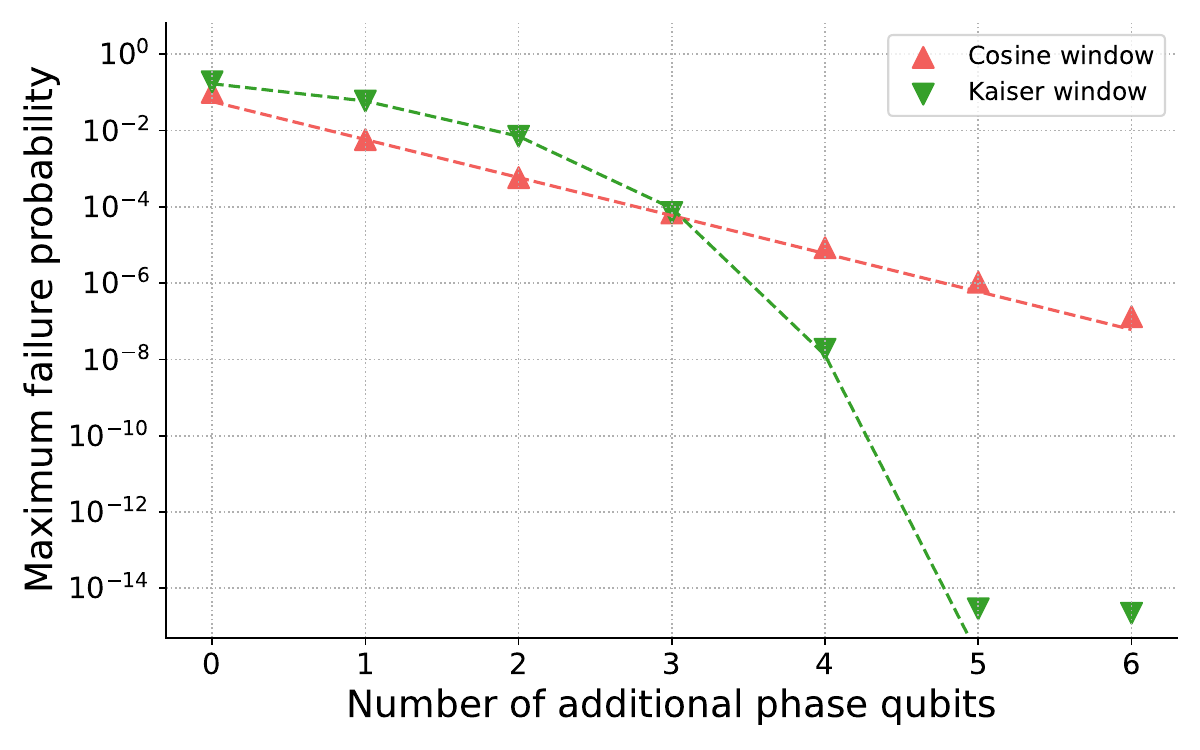}
\caption{Maximum QPE failure probability (defined as $1-P_{\text{succ}}$) as a function of the number of additional phase qubits that are added and then discarded to obtain a more accurate estimate of the eigenphase using the cosine (red triangles) and Kaiser (light blue inverted triangles) window functions. $P_{\text{succ}}$ is the minimum success probability.
Both show an exponential scaling from one additional phase qubit onwards, although the Kaiser window performs significantly better. For 5 and 6 additional phase qubits, the Kaiser window failure probabilities are at the threshold for floating point error and so those values should not be taken as being representative of the QPE performance with those numbers of qubits. The estimates of the eigenphase were made using 5 bits of precision. 
The fitted lines show the expected $\log(1/\delta)$ and $\log \log(1/\delta)$ scalings for the cosine and Kaiser windows respectively (see Table~\ref{tab:window_functions}).
}
\label{fig:window_func_extra_qubits}
\end{figure}

As a first step towards answering this question, we evaluate the impact of increasing the number of additional phase qubits (that are subsequently discarded) for the two additional window functions considered here, the cosine and the Kaiser windows. Fig.~\ref{fig:window_func_extra_qubits} shows the maximum failure probability for these two windows as a function of the number of additional phase qubits. The cosine window shows an exponential improvement in failure probability with additional phase qubits, reaching a minimum failure probability of $10^{-7}$ using 6 additional phase qubits, while the Kaiser window shows an exponential improvement {\it over that}, with the obtained failure probabilities being limited by floating point error after only 4 additional phase qubits. This performance closely matches the expected asymptotic performance, as shown by the fits to the expected $\log(1/\delta)$ and $\log \log(1/\delta)$ scalings for the cosine and Kaiser windows respectively that are shown in Fig.~\ref{fig:window_func_extra_qubits}.

It is worth highlighting just how overwhelmingly more performant the window functions are over QSVT QPE: 6 additional phase qubits is only slightly more expensive than performing a 64 degree polynomial approximation in QSVT QPE, which yields a maximum failure probability of approximately $10^{-2}$ as shown in Fig~\ref{fig:success_probs_5_bits} compared with $10^{-7}$ for the cosine window. For the Kaiser window, the limitations of double precision floating point arithmetic mean that the largest number of additional phase qubits with reliable success probabilities is 4, corresponding to a QPE routine with a query cost approximately 1/4 that of the 64-degree polynomial. Despite this, the achieved failure probability is less than $10^{-7}$, some five orders of magnitude less than the QSVT QPE results. One may argue that QSVT QPE may be more advantageous at extremely low failure probabilities (for example that resulting from the $10^{-30}$ target polynomial error used in Ref.~\cite{rall2021faster}). However, we argue that such a cross-over is unlikely given the numerical evidence presented here, and that even if it does occur, it would be well beyond the success probabilities needed for practical applications.

The data in Fig.~\ref{fig:window_func_extra_qubits} were obtained by evaluating the failure probability across the periodic range of eigenphases in $[0, \frac{2\pi}{2^{b}}]$ and taking the maximal failure probability. These data therefore represent the {\it worst-case} failure probability and therefore one would expect to do even better than this in the typical case (although one cannot predict where the true eigenphase lies, and so this worst-case bound is the most appropriate figure of merit for evaluating the viability of algorithms). It should also be noted that for QSVT QPE, the assumption that the failure probabilities are periodic is not valid due to the choice of polynomial approximation, with this choice of eigenphases being the worst-case choice for QSVT QPE – Appendix~\ref{app:qsvt_succ_prob_assymetry} outlines the reason for the breaking of this symmetry.

\subsection{Numerical success probability scaling with bits of precision}

In the previous sections, we presented numerical evidence that the success probability of QPE can be increased by either using window states or by using QSVT as outlined in Ref.~\cite{martyn2021grand}, 
arguing that window functions use far fewer resources to accomplish similar increases in success probability over the standard QPE routine. 
However, a natural question arises as to whether these results are reliable for all domains of interest -- the number of bits of precision required to estimate an eigenphase to a given precision varies significantly (e.g. in quantum chemistry applications, it scales with the norm of the system Hamiltonian) and in practice can be large enough to make classical simulation impractical. 
In this section, we present numerical evidence that allows us to make a heuristic argument about the performance of these strategies for larger numbers of phase qubits.

\begin{figure}[h]
\centering
\includegraphics[width=0.6\textwidth]{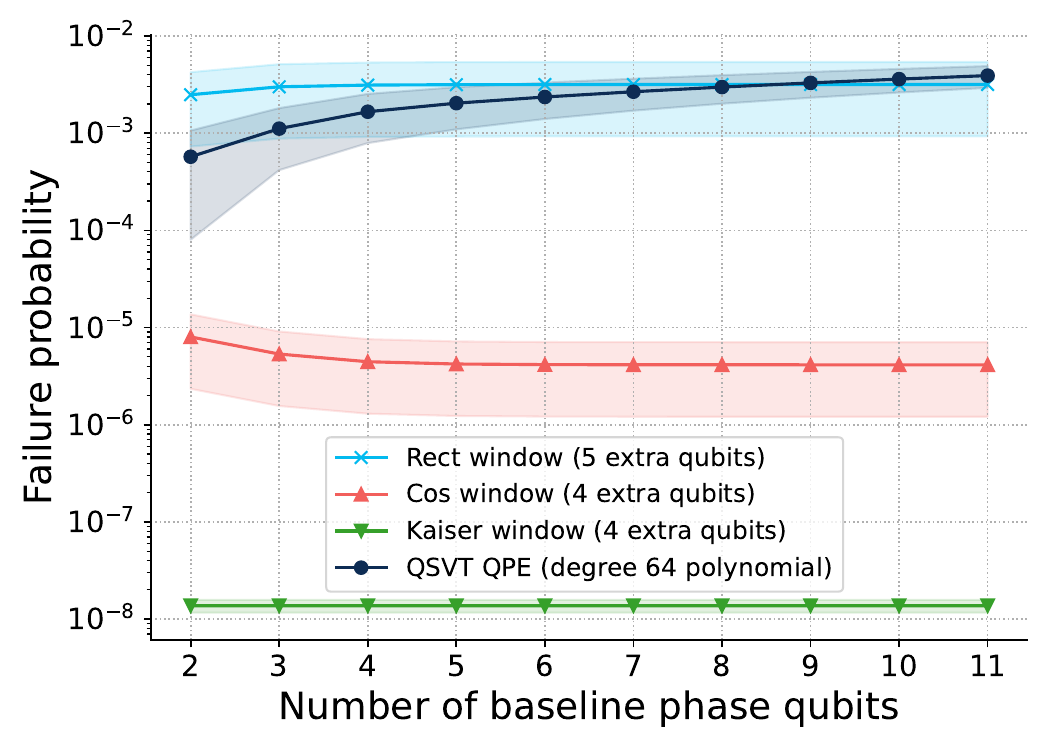}
\caption{QPE success probability for the three different window states (rectangular, cosine and Kaiser) and QSVT QPE as a function of the number of bits of precision in the phase register. The points indicate the mean of the success probabilities evaluated over $10 000$ points evenly distributed in $[0, \frac{2\pi}{2^b}]$ and the shaded regions correspond to the standard deviations of the same data. For all three window states, both the mean and the standard deviations remain constant over all bits of precision considered, providing solid evidence that they will remain highly effective even at the larger register sizes required for practical applications. QSVT QPE shows a linear decrease in mean success probability.}
\label{fig:success_prob_scaling}
\end{figure}

We evaluate the scaling of the different QPE methods by numerically generating the success probability for different values of eigenphases for $m$ bits of precision from 1 to 14, once again using the phase unitary as the test bed. Although this is likely fewer bits of precision than would be required to estimate, for example, the ground state of FeMocco to chemical accuracy~\cite{lee2021even}, we argue that there is no a priori reason to believe that the scaling observed for these parameters should break down at higher system sizes. For the window functions, an efficient emulation was implemented that allowed us to sample 10~000 points along a range of eigenphases corresponding to a period of success probabilities, or $[0, 1/2^{n}]$ for $n = m + p$ total phase qubits, while for QSVT QPE a full statevector simulation of the circuit using 10~000 points over the full domain $[0, 2\pi]$. The reason for this discrepancy is that the periodicity of the QPE routine is broken by the imperfect polynomial approximation (see Appendix~\ref{app:qsvt_succ_prob_assymetry} for details), meaning that unlike the window functions, sampling over a single period does not result in comparable success probabilities to sampling over the entire domain. Fig.~\ref{fig:success_prob_scaling} shows the results of this numerical evaluation. The performance of the window functions is independent of the number of bits of precision required, with both the mean and standard deviations remaining constant for all simulations. This provides a significant boost to the utility of these functions: if one needs some guarantee on the accuracy of their QPE routine, then they can determine how many additional qubits will be required using only a small numerical model.

For QSVT QPE, the success probability decreases as the number of bits of precision increases, although it remains high for the system sizes considered here. This decrease means that the success probabilities obtained in the data presented in the main text will not be representative of the performance at large numbers of bits of precision, but rather form an upper bound on the performance. That being said, a reasonable guess as to the required degree can be made by extrapolating the plot out to larger system sizes from data such as this.

The reason that the success probability of QSVT-QPE decreases as a function of the number of bits of precision is due to the fact that the number of applications of the function (and the exponentiated unitaries) increase as a function of the number of bits. Each QSVT application remains in the desired eigenspace with probability approximately equal to $1-\Delta$ (ignoring the region of discontinuity around phase values close to $1/\sqrt{2}$, which will further decrease the success probability), meaning that for $m$ bits of precision, the overall probability of successfully implementing the desired function, rather than projecting into some orthogonal eigenspace, is $1 - \delta \approx (1-\Delta)^m$. Thus, for any finite degree approximation to the sign function, the success probability decreases exponentially in the number of phase bits. By contrast, the window functions implement a well-defined state on the phase register regardless of the number of bits of precision, and therefore the performance of the window functions is independent of the number of phase bits (although the resources required to implement the window function will of course depend on the size of the phase register, and in the worst case exponentially – this is relatively unproblematic however, since the number of block encoding applications also scales exponentially with the number of bits of precision, and the cost of even a single block encoding will typically dwarf the cost of implementing the window functions). Given that the main results presented here show that window functions achieve significantly better success probabilities with lower costs than QSVT QPE, this provides further evidence that QSVT is not an effective strategy for increasing the performance of QPE.

\section{Conclusions and outlook}

In this work we compare the effectiveness of two approaches towards increasing the success probability of QPE, a property that will need to be maximized in order for coherent usage of the subroutine to be possible.\\

\noindent \textbf{Utility of QSVT QPE and window functions}
Our numerical results show that while QSVT QPE performs as well as expected from previous asymptotic analysis, it is significantly outperformed by using window functions together with additional phase qubits. QSVT QPE using a degree 64 polynomial yielded a slightly higher minimum success probability than the rectangular window function with 5 additional phase qubits, but required almost twice the number of calls to the block encoding. The performance gap is even wider when comparing QSVT QPE to the cosine and Kaiser window functions, which achieve substantially higher success probabilities for 1/4 the query cost. The Kaiser window shows a particularly overwhelming improvement in performance compared with QSVT QPE, with the failure probability hitting the precision floor for double precision floating point numbers with only $m=5$ additional phase bits, still half the query cost QSVT QPE requires to achieve a success probability of 0.99.

These results indicate that QPE is not an algorithm which benefits significantly from employing QSVT as a subroutine. It is possible that other attempts to combine QSVT and QPE could have some advantage -- the implementation presented in Ref.~\cite{rall2021faster} has an improved asymptotic scaling, for example. However, given the performance of the Kaiser window and the extremely high success probabilities it was able to achieve with fewer than 6 additional phase qubits, the likelihood is that for most applications, the Kaiser window will be more than sufficient for implementing QPE with high success probabilities.\\

\noindent \textbf{Future directions}
There are many future directions that could be taken to expand upon the work we present here. One idea would be to combine QSVT QPE with the window functions in order to potentially improve the performance of both. Using the QSVT QPE framework as is, we do not expect the two methods to be compatible as QSVT-QPE uses a thresholding function to reduce ambiguity in determining each bit value. This corresponds to the bit sitting exactly on top of a bin, a case in which using the rectangular window function is optimal.
We attempted a simple implementation of this by taking the QSVT QPE routine and using the cosine window in place of the rectangular window, which resulted in a marked decrease in performance compared with the standard QSVT QPE routine. 
It may be possible that some modification to the QSVT QPE routine could be made that directly accounts for the different window functions. 
The viability of such an approach would be an interesting study for future work. 
Additionally,
it would be worth verifying whether the alternative implementations of QSVT QPE mentioned above can outperform the implementation investigated here.
However, given the remarkable performance of the Kaiser window, in which the failure probability using four extra bits already approaches the floating point error, it seems unlikely that any implementation would be more efficient than that. This would require the QSVT QPE to use a 16-degree polynomial for the same task to be competitive with the use of the Kaiser window.

The state vector simulations using the Kaiser window were not implemented with a full gate decomposition -- rather they were implemented by directly injecting the desired state into the state vector before applying the QPE routine. A cost analysis for the implementation of the Kaiser window is given in Appendix~\ref{sec:envelope}, but this is merely a worst-case upper bound on the cost, assuming that arbitrary state preparation is required. Given the significant amount of structure in the definition of the Kaiser window and the fact it can be exactly expressed in terms of Bessel functions, it is likely that more efficient implementations can be obtained. It would also be interesting to analyze the impact of implementing an approximation to the Kaiser window, which may allow for more favorable balances between cost and success probability.

In our simulations, we also assume that we are exactly implementing the input eigenstate such that the only error in the phase estimation arises from the finite bit precision of the phase register. In practise, such an assumption will not hold for real systems. It would therefore be instructive to investigate the properties of the different approaches when the input eigenstate is only approximately realized.

As a more general point, our results have highlighted a limitation with the applicability of QSVT in QPE. As a framework, QSVT is highly general, and can be applied to a wide variety of applications. It would be highly valuable to investigate whether other proposed algorithms, such as Ref.~\cite{wang2023quantum}, making use of QSP or QSVT have similar limitations, or whether they can genuinely be relied upon to yield efficient quantum algorithms. This is especially true in light of the recent publication of generalized quantum signal processing~\cite{motlagh2023generalized} which relaxes many of the constraints on the target functions of QSVT and hence makes the technique more appealing for applications that can take advantage of this improvement. The target function used in QSVT QPE, however, already naturally accommodates these restrictions and so QSVT QPE would not benefit from this generalization. \\

\noindent \textbf{Concluding remarks}
In this work we present a systematic numerical comparison of two different methods for increasing the success probability of QPE, one using window functions from signal processing and one using the quantum singular value transform (QSVT QPE). From these numerical results, using a Kaiser window appears to be the best method for improving the success probability of QPE, providing the greatest increase for the lowest cost. Such a conclusion would be difficult to ascertain from an analysis of the asymptotic scaling alone, highlighting the importance of concrete verification protocols in quantum computation. In addition to the practical utility for quantum algorithms designers our work entails, we hope this study spurs further work into both numerical verification procedures and the systematic evaluation of existing protocols in quantum computation.

\section*{Author contributions}
SS proposed comparison of window functions and QSVT for QPE. SG and SS developed codes for QSVT and window functions, respectively. SG developed and ran simulations for QPE using QSVT and both SG and SS for window functions. WP approximated the cost for preparing the Kaiser window state. SG, WP, and SS discussed the project and wrote the manuscript.

\section*{Acknowledgement}
The authors would like to thank Eric Johnston for his support in developing and debugging our simulation code. We would also like to thank Mark Steudtner, Jessica Lemieux, Harriet Apel, and Sam Pallister for insightful discussions on QPE.

\bibliographystyle{apsrev4-1}
\bibliography{bibliography}

\begin{thebibliography}{50}%
\makeatletter
\providecommand \@ifxundefined [1]{%
 \@ifx{#1\undefined}
}%
\providecommand \@ifnum [1]{%
 \ifnum #1\expandafter \@firstoftwo
 \else \expandafter \@secondoftwo
 \fi
}%
\providecommand \@ifx [1]{%
 \ifx #1\expandafter \@firstoftwo
 \else \expandafter \@secondoftwo
 \fi
}%
\providecommand \natexlab [1]{#1}%
\providecommand \enquote  [1]{``#1''}%
\providecommand \bibnamefont  [1]{#1}%
\providecommand \bibfnamefont [1]{#1}%
\providecommand \citenamefont [1]{#1}%
\providecommand \href@noop [0]{\@secondoftwo}%
\providecommand \href [0]{\begingroup \@sanitize@url \@href}%
\providecommand \@href[1]{\@@startlink{#1}\@@href}%
\providecommand \@@href[1]{\endgroup#1\@@endlink}%
\providecommand \@sanitize@url [0]{\catcode `\\12\catcode `\$12\catcode
  `\&12\catcode `\#12\catcode `\^12\catcode `\_12\catcode `\%12\relax}%
\providecommand \@@startlink[1]{}%
\providecommand \@@endlink[0]{}%
\providecommand \url  [0]{\begingroup\@sanitize@url \@url }%
\providecommand \@url [1]{\endgroup\@href {#1}{\urlprefix }}%
\providecommand \urlprefix  [0]{URL }%
\providecommand \Eprint [0]{\href }%
\providecommand \doibase [0]{http://dx.doi.org/}%
\providecommand \selectlanguage [0]{\@gobble}%
\providecommand \bibinfo  [0]{\@secondoftwo}%
\providecommand \bibfield  [0]{\@secondoftwo}%
\providecommand \translation [1]{[#1]}%
\providecommand \BibitemOpen [0]{}%
\providecommand \bibitemStop [0]{}%
\providecommand \bibitemNoStop [0]{.\EOS\space}%
\providecommand \EOS [0]{\spacefactor3000\relax}%
\providecommand \BibitemShut  [1]{\csname bibitem#1\endcsname}%
\let\auto@bib@innerbib\@empty
\bibitem [{\citenamefont {Grover}(1996)}]{grover1996fast}%
  \BibitemOpen
  \bibfield  {author} {\bibinfo {author} {\bibfnamefont {L.~K.}\ \bibnamefont
  {Grover}},\ }in\ \href {\doibase 10.1145/237814.237866} {\emph {\bibinfo
  {booktitle} {Proceedings of the Twenty-Eighth Annual ACM Symposium on Theory
  of Computing}}},\ \bibinfo {series and number} {STOC '96}\ (\bibinfo
  {publisher} {Association for Computing Machinery},\ \bibinfo {address} {New
  York, NY, USA},\ \bibinfo {year} {1996})\ p.\ \bibinfo {pages}
  {212–219}\BibitemShut {NoStop}%
\bibitem [{\citenamefont {Feynman}(1982)}]{feynman1982simulating}%
  \BibitemOpen
  \bibfield  {author} {\bibinfo {author} {\bibfnamefont {R.~P.}\ \bibnamefont
  {Feynman}},\ }\href {\doibase 10.1007/BF02650179} {\bibfield  {journal}
  {\bibinfo  {journal} {Int. J. Theor. Phys.}\ }\textbf {\bibinfo {volume}
  {21}},\ \bibinfo {pages} {467–488} (\bibinfo {year} {1982})}\BibitemShut
  {NoStop}%
\bibitem [{\citenamefont {Georgescu}\ \emph {et~al.}(2014)\citenamefont
  {Georgescu}, \citenamefont {Ashhab},\ and\ \citenamefont
  {Nori}}]{georgescu2014quantum}%
  \BibitemOpen
  \bibfield  {author} {\bibinfo {author} {\bibfnamefont {I.~M.}\ \bibnamefont
  {Georgescu}}, \bibinfo {author} {\bibfnamefont {S.}~\bibnamefont {Ashhab}}, \
  and\ \bibinfo {author} {\bibfnamefont {F.}~\bibnamefont {Nori}},\ }\href
  {\doibase 10.1103/RevModPhys.86.153} {\bibfield  {journal} {\bibinfo
  {journal} {Rev. Mod. Phys.}\ }\textbf {\bibinfo {volume} {86}},\ \bibinfo
  {pages} {153} (\bibinfo {year} {2014})}\BibitemShut {NoStop}%
\bibitem [{\citenamefont {Shor}(1994)}]{shor1994algorithms}%
  \BibitemOpen
  \bibfield  {author} {\bibinfo {author} {\bibfnamefont {P.}~\bibnamefont
  {Shor}},\ }in\ \href {\doibase 10.1109/SFCS.1994.365700} {\emph {\bibinfo
  {booktitle} {Proceedings 35th Annual Symposium on Foundations of Computer
  Science}}}\ (\bibinfo {year} {1994})\ pp.\ \bibinfo {pages}
  {124--134}\BibitemShut {NoStop}%
\bibitem [{\citenamefont {Shor}(1997)}]{shor1997polynomial}%
  \BibitemOpen
  \bibfield  {author} {\bibinfo {author} {\bibfnamefont {P.~W.}\ \bibnamefont
  {Shor}},\ }\href {\doibase 10.1137/S0097539795293172} {\bibfield  {journal}
  {\bibinfo  {journal} {SIAM Journal on Computing}\ }\textbf {\bibinfo {volume}
  {26}},\ \bibinfo {pages} {1484} (\bibinfo {year} {1997})},\ \Eprint
  {http://arxiv.org/abs/https://doi.org/10.1137/S0097539795293172}
  {https://doi.org/10.1137/S0097539795293172} \BibitemShut {NoStop}%
\bibitem [{\citenamefont {Kitaev}(1995)}]{kitaev1995quantum}%
  \BibitemOpen
  \bibfield  {author} {\bibinfo {author} {\bibfnamefont {A.~Y.}\ \bibnamefont
  {Kitaev}},\ }\href@noop {} {\enquote {\bibinfo {title} {Quantum measurements
  and the abelian stabilizer problem},}\ } (\bibinfo {year} {1995}),\ \Eprint
  {http://arxiv.org/abs/quant-ph/9511026} {arXiv:quant-ph/9511026 [quant-ph]}
  \BibitemShut {NoStop}%
\bibitem [{\citenamefont {Abrams}\ and\ \citenamefont
  {Lloyd}(1999)}]{abrams1999quantum}%
  \BibitemOpen
  \bibfield  {author} {\bibinfo {author} {\bibfnamefont {D.~S.}\ \bibnamefont
  {Abrams}}\ and\ \bibinfo {author} {\bibfnamefont {S.}~\bibnamefont {Lloyd}},\
  }\href {\doibase 10.1103/PhysRevLett.83.5162} {\bibfield  {journal} {\bibinfo
   {journal} {Phys. Rev. Lett.}\ }\textbf {\bibinfo {volume} {83}},\ \bibinfo
  {pages} {5162} (\bibinfo {year} {1999})}\BibitemShut {NoStop}%
\bibitem [{\citenamefont {Gidney}(2018{\natexlab{a}})}]{gidney2018factoring}%
  \BibitemOpen
  \bibfield  {author} {\bibinfo {author} {\bibfnamefont {C.}~\bibnamefont
  {Gidney}},\ }\href@noop {} {\enquote {\bibinfo {title} {Factoring with n+2
  clean qubits and n-1 dirty qubits},}\ } (\bibinfo {year}
  {2018}{\natexlab{a}}),\ \Eprint {http://arxiv.org/abs/1706.07884}
  {arXiv:1706.07884 [quant-ph]} \BibitemShut {NoStop}%
\bibitem [{\citenamefont {Litinski}(2023)}]{litinski2023compute}%
  \BibitemOpen
  \bibfield  {author} {\bibinfo {author} {\bibfnamefont {D.}~\bibnamefont
  {Litinski}},\ }\href@noop {} {\enquote {\bibinfo {title} {How to compute a
  256-bit elliptic curve private key with only 50 million toffoli gates},}\ }
  (\bibinfo {year} {2023}),\ \Eprint {http://arxiv.org/abs/2306.08585}
  {arXiv:2306.08585 [quant-ph]} \BibitemShut {NoStop}%
\bibitem [{\citenamefont {Whitfield}\ \emph {et~al.}(2011)\citenamefont
  {Whitfield}, \citenamefont {Biamonte},\ and\ \citenamefont
  {Aspuru-Guzik}}]{whitfield2011simulation}%
  \BibitemOpen
  \bibfield  {author} {\bibinfo {author} {\bibfnamefont {J.~D.}\ \bibnamefont
  {Whitfield}}, \bibinfo {author} {\bibfnamefont {J.}~\bibnamefont {Biamonte}},
  \ and\ \bibinfo {author} {\bibfnamefont {A.}~\bibnamefont {Aspuru-Guzik}},\
  }\href {\doibase 10.1080/00268976.2011.552441} {\bibfield  {journal}
  {\bibinfo  {journal} {Molecular Physics}\ }\textbf {\bibinfo {volume}
  {109}},\ \bibinfo {pages} {735} (\bibinfo {year} {2011})},\ \Eprint
  {http://arxiv.org/abs/https://doi.org/10.1080/00268976.2011.552441}
  {https://doi.org/10.1080/00268976.2011.552441} \BibitemShut {NoStop}%
\bibitem [{\citenamefont {Reiher}\ \emph {et~al.}(2017)\citenamefont {Reiher},
  \citenamefont {Wiebe}, \citenamefont {Svore}, \citenamefont {Wecker},\ and\
  \citenamefont {Troyer}}]{reiher2017elucidating}%
  \BibitemOpen
  \bibfield  {author} {\bibinfo {author} {\bibfnamefont {M.}~\bibnamefont
  {Reiher}}, \bibinfo {author} {\bibfnamefont {N.}~\bibnamefont {Wiebe}},
  \bibinfo {author} {\bibfnamefont {K.~M.}\ \bibnamefont {Svore}}, \bibinfo
  {author} {\bibfnamefont {D.}~\bibnamefont {Wecker}}, \ and\ \bibinfo {author}
  {\bibfnamefont {M.}~\bibnamefont {Troyer}},\ }\href {\doibase
  10.1073/pnas.1619152114} {\bibfield  {journal} {\bibinfo  {journal}
  {Proceedings of the National Academy of Sciences}\ }\textbf {\bibinfo
  {volume} {114}},\ \bibinfo {pages} {7555} (\bibinfo {year} {2017})},\ \Eprint
  {http://arxiv.org/abs/https://www.pnas.org/doi/pdf/10.1073/pnas.1619152114}
  {https://www.pnas.org/doi/pdf/10.1073/pnas.1619152114} \BibitemShut {NoStop}%
\bibitem [{\citenamefont {Babbush}\ \emph
  {et~al.}(2018{\natexlab{a}})\citenamefont {Babbush}, \citenamefont {Wiebe},
  \citenamefont {McClean}, \citenamefont {McClain}, \citenamefont {Neven},\
  and\ \citenamefont {Chan}}]{babbush2018low}%
  \BibitemOpen
  \bibfield  {author} {\bibinfo {author} {\bibfnamefont {R.}~\bibnamefont
  {Babbush}}, \bibinfo {author} {\bibfnamefont {N.}~\bibnamefont {Wiebe}},
  \bibinfo {author} {\bibfnamefont {J.}~\bibnamefont {McClean}}, \bibinfo
  {author} {\bibfnamefont {J.}~\bibnamefont {McClain}}, \bibinfo {author}
  {\bibfnamefont {H.}~\bibnamefont {Neven}}, \ and\ \bibinfo {author}
  {\bibfnamefont {G.~K.-L.}\ \bibnamefont {Chan}},\ }\href {\doibase
  10.1103/PhysRevX.8.011044} {\bibfield  {journal} {\bibinfo  {journal} {Phys.
  Rev. X}\ }\textbf {\bibinfo {volume} {8}},\ \bibinfo {pages} {011044}
  (\bibinfo {year} {2018}{\natexlab{a}})}\BibitemShut {NoStop}%
\bibitem [{\citenamefont {Babbush}\ \emph
  {et~al.}(2018{\natexlab{b}})\citenamefont {Babbush}, \citenamefont {Gidney},
  \citenamefont {Berry}, \citenamefont {Wiebe}, \citenamefont {McClean},
  \citenamefont {Paler}, \citenamefont {Fowler},\ and\ \citenamefont
  {Neven}}]{babbush2018encoding}%
  \BibitemOpen
  \bibfield  {author} {\bibinfo {author} {\bibfnamefont {R.}~\bibnamefont
  {Babbush}}, \bibinfo {author} {\bibfnamefont {C.}~\bibnamefont {Gidney}},
  \bibinfo {author} {\bibfnamefont {D.~W.}\ \bibnamefont {Berry}}, \bibinfo
  {author} {\bibfnamefont {N.}~\bibnamefont {Wiebe}}, \bibinfo {author}
  {\bibfnamefont {J.}~\bibnamefont {McClean}}, \bibinfo {author} {\bibfnamefont
  {A.}~\bibnamefont {Paler}}, \bibinfo {author} {\bibfnamefont
  {A.}~\bibnamefont {Fowler}}, \ and\ \bibinfo {author} {\bibfnamefont
  {H.}~\bibnamefont {Neven}},\ }\href {\doibase 10.1103/physrevx.8.041015}
  {\bibfield  {journal} {\bibinfo  {journal} {Physical Review X}\ }\textbf
  {\bibinfo {volume} {8}},\ \bibinfo {pages} {041015} (\bibinfo {year}
  {2018}{\natexlab{b}})},\ \Eprint {http://arxiv.org/abs/1805.03662}
  {1805.03662} \BibitemShut {NoStop}%
\bibitem [{\citenamefont {Kivlichan}\ \emph {et~al.}(2020)\citenamefont
  {Kivlichan}, \citenamefont {Gidney}, \citenamefont {Berry}, \citenamefont
  {Wiebe}, \citenamefont {McClean}, \citenamefont {Sun}, \citenamefont {Jiang},
  \citenamefont {Rubin}, \citenamefont {Fowler}, \citenamefont {Aspuru-Guzik},
  \citenamefont {Neven},\ and\ \citenamefont
  {Babbush}}]{Kivlichan2020improvedfault}%
  \BibitemOpen
  \bibfield  {author} {\bibinfo {author} {\bibfnamefont {I.~D.}\ \bibnamefont
  {Kivlichan}}, \bibinfo {author} {\bibfnamefont {C.}~\bibnamefont {Gidney}},
  \bibinfo {author} {\bibfnamefont {D.~W.}\ \bibnamefont {Berry}}, \bibinfo
  {author} {\bibfnamefont {N.}~\bibnamefont {Wiebe}}, \bibinfo {author}
  {\bibfnamefont {J.}~\bibnamefont {McClean}}, \bibinfo {author} {\bibfnamefont
  {W.}~\bibnamefont {Sun}}, \bibinfo {author} {\bibfnamefont {Z.}~\bibnamefont
  {Jiang}}, \bibinfo {author} {\bibfnamefont {N.}~\bibnamefont {Rubin}},
  \bibinfo {author} {\bibfnamefont {A.}~\bibnamefont {Fowler}}, \bibinfo
  {author} {\bibfnamefont {A.}~\bibnamefont {Aspuru-Guzik}}, \bibinfo {author}
  {\bibfnamefont {H.}~\bibnamefont {Neven}}, \ and\ \bibinfo {author}
  {\bibfnamefont {R.}~\bibnamefont {Babbush}},\ }\href {\doibase
  10.22331/q-2020-07-16-296} {\bibfield  {journal} {\bibinfo  {journal}
  {{Quantum}}\ }\textbf {\bibinfo {volume} {4}},\ \bibinfo {pages} {296}
  (\bibinfo {year} {2020})}\BibitemShut {NoStop}%
\bibitem [{\citenamefont {Lee}\ \emph {et~al.}(2021)\citenamefont {Lee},
  \citenamefont {Berry}, \citenamefont {Gidney}, \citenamefont {Huggins},
  \citenamefont {McClean}, \citenamefont {Wiebe},\ and\ \citenamefont
  {Babbush}}]{lee2021even}%
  \BibitemOpen
  \bibfield  {author} {\bibinfo {author} {\bibfnamefont {J.}~\bibnamefont
  {Lee}}, \bibinfo {author} {\bibfnamefont {D.~W.}\ \bibnamefont {Berry}},
  \bibinfo {author} {\bibfnamefont {C.}~\bibnamefont {Gidney}}, \bibinfo
  {author} {\bibfnamefont {W.~J.}\ \bibnamefont {Huggins}}, \bibinfo {author}
  {\bibfnamefont {J.~R.}\ \bibnamefont {McClean}}, \bibinfo {author}
  {\bibfnamefont {N.}~\bibnamefont {Wiebe}}, \ and\ \bibinfo {author}
  {\bibfnamefont {R.}~\bibnamefont {Babbush}},\ }\href {\doibase
  10.1103/PRXQuantum.2.030305} {\bibfield  {journal} {\bibinfo  {journal} {PRX
  Quantum}\ }\textbf {\bibinfo {volume} {2}},\ \bibinfo {pages} {030305}
  (\bibinfo {year} {2021})}\BibitemShut {NoStop}%
\bibitem [{\citenamefont {Gilyen}\ \emph {et~al.}(2019)\citenamefont {Gilyen},
  \citenamefont {Su}, \citenamefont {Low},\ and\ \citenamefont
  {Wiebe}}]{gilyen2019quantum}%
  \BibitemOpen
  \bibfield  {author} {\bibinfo {author} {\bibfnamefont {A.}~\bibnamefont
  {Gilyen}}, \bibinfo {author} {\bibfnamefont {Y.}~\bibnamefont {Su}}, \bibinfo
  {author} {\bibfnamefont {G.~H.}\ \bibnamefont {Low}}, \ and\ \bibinfo
  {author} {\bibfnamefont {N.}~\bibnamefont {Wiebe}},\ }\href {\doibase
  10.1145/3313276.3316366} {\bibfield  {journal} {\bibinfo  {journal}
  {Proceedings of the 51st Annual ACM SIGACT Symposium on Theory of Computing}\
  ,\ \bibinfo {pages} {193}} (\bibinfo {year} {2019})},\ \Eprint
  {http://arxiv.org/abs/1806.01838} {1806.01838} \BibitemShut {NoStop}%
\bibitem [{\citenamefont {Rall}(2021)}]{rall2021faster}%
  \BibitemOpen
  \bibfield  {author} {\bibinfo {author} {\bibfnamefont {P.}~\bibnamefont
  {Rall}},\ }\href {\doibase 10.22331/q-2021-10-19-566} {\bibfield  {journal}
  {\bibinfo  {journal} {Quantum}\ }\textbf {\bibinfo {volume} {5}},\ \bibinfo
  {pages} {566} (\bibinfo {year} {2021})},\ \Eprint
  {http://arxiv.org/abs/2103.09717} {2103.09717} \BibitemShut {NoStop}%
\bibitem [{\citenamefont {Martyn}\ \emph {et~al.}(2021)\citenamefont {Martyn},
  \citenamefont {Rossi}, \citenamefont {Tan},\ and\ \citenamefont
  {Chuang}}]{martyn2021grand}%
  \BibitemOpen
  \bibfield  {author} {\bibinfo {author} {\bibfnamefont {J.~M.}\ \bibnamefont
  {Martyn}}, \bibinfo {author} {\bibfnamefont {Z.~M.}\ \bibnamefont {Rossi}},
  \bibinfo {author} {\bibfnamefont {A.~K.}\ \bibnamefont {Tan}}, \ and\
  \bibinfo {author} {\bibfnamefont {I.~L.}\ \bibnamefont {Chuang}},\ }\href
  {\doibase 10.1103/prxquantum.2.040203} {\bibfield  {journal} {\bibinfo
  {journal} {PRX Quantum}\ }\textbf {\bibinfo {volume} {2}},\ \bibinfo {pages}
  {040203} (\bibinfo {year} {2021})},\ \Eprint
  {http://arxiv.org/abs/2105.02859} {2105.02859} \BibitemShut {NoStop}%
\bibitem [{\citenamefont {Rendon}\ \emph {et~al.}(2022)\citenamefont {Rendon},
  \citenamefont {Izubuchi},\ and\ \citenamefont {Kikuchi}}]{rendon2022effects}%
  \BibitemOpen
  \bibfield  {author} {\bibinfo {author} {\bibfnamefont {G.}~\bibnamefont
  {Rendon}}, \bibinfo {author} {\bibfnamefont {T.}~\bibnamefont {Izubuchi}}, \
  and\ \bibinfo {author} {\bibfnamefont {Y.}~\bibnamefont {Kikuchi}},\ }\href
  {\doibase 10.1103/physrevd.106.034503} {\bibfield  {journal} {\bibinfo
  {journal} {Physical Review D}\ }\textbf {\bibinfo {volume} {106}},\ \bibinfo
  {pages} {034503} (\bibinfo {year} {2022})},\ \Eprint
  {http://arxiv.org/abs/2110.09590} {2110.09590} \BibitemShut {NoStop}%
\bibitem [{\citenamefont {Patel}\ \emph {et~al.}(2024)\citenamefont {Patel},
  \citenamefont {Tan}, \citenamefont {Suba\c{s}\i},\ and\ \citenamefont
  {Sornborger}}]{patel2024optimal}%
  \BibitemOpen
  \bibfield  {author} {\bibinfo {author} {\bibfnamefont {D.}~\bibnamefont
  {Patel}}, \bibinfo {author} {\bibfnamefont {S.~J.~S.}\ \bibnamefont {Tan}},
  \bibinfo {author} {\bibfnamefont {Y.}~\bibnamefont {Suba\c{s}\i}}, \ and\
  \bibinfo {author} {\bibfnamefont {A.~T.}\ \bibnamefont {Sornborger}},\
  }\href@noop {} {\bibfield  {journal} {\bibinfo  {journal} {arXiv}\ }
  (\bibinfo {year} {2024})},\ \Eprint {http://arxiv.org/abs/2403.18927}
  {2403.18927} \BibitemShut {NoStop}%
\bibitem [{\citenamefont {Harrow}\ \emph {et~al.}(2009)\citenamefont {Harrow},
  \citenamefont {Hassidim},\ and\ \citenamefont {Lloyd}}]{hhl}%
  \BibitemOpen
  \bibfield  {author} {\bibinfo {author} {\bibfnamefont {A.~W.}\ \bibnamefont
  {Harrow}}, \bibinfo {author} {\bibfnamefont {A.}~\bibnamefont {Hassidim}}, \
  and\ \bibinfo {author} {\bibfnamefont {S.}~\bibnamefont {Lloyd}},\ }\href
  {\doibase 10.1103/PhysRevLett.103.150502} {\bibfield  {journal} {\bibinfo
  {journal} {Phys. Rev. Lett.}\ }\textbf {\bibinfo {volume} {103}},\ \bibinfo
  {pages} {150502} (\bibinfo {year} {2009})}\BibitemShut {NoStop}%
\bibitem [{\citenamefont {Temme}\ \emph {et~al.}(2011)\citenamefont {Temme},
  \citenamefont {Osborne}, \citenamefont {Vollbrecht}, \citenamefont {Poulin},\
  and\ \citenamefont {Verstraete}}]{metropolis_sampling}%
  \BibitemOpen
  \bibfield  {author} {\bibinfo {author} {\bibfnamefont {K.}~\bibnamefont
  {Temme}}, \bibinfo {author} {\bibfnamefont {T.~J.}\ \bibnamefont {Osborne}},
  \bibinfo {author} {\bibfnamefont {K.~G.}\ \bibnamefont {Vollbrecht}},
  \bibinfo {author} {\bibfnamefont {D.}~\bibnamefont {Poulin}}, \ and\ \bibinfo
  {author} {\bibfnamefont {F.}~\bibnamefont {Verstraete}},\ }\href {\doibase
  10.1038/nature09770} {\bibfield  {journal} {\bibinfo  {journal} {Nature}\
  }\textbf {\bibinfo {volume} {471}},\ \bibinfo {pages} {87–90} (\bibinfo
  {year} {2011})}\BibitemShut {NoStop}%
\bibitem [{\citenamefont {Yung}\ and\ \citenamefont
  {Aspuru-Guzik}(2012)}]{yung2012quantum}%
  \BibitemOpen
  \bibfield  {author} {\bibinfo {author} {\bibfnamefont {M.-H.}\ \bibnamefont
  {Yung}}\ and\ \bibinfo {author} {\bibfnamefont {A.}~\bibnamefont
  {Aspuru-Guzik}},\ }\href@noop {} {\bibfield  {journal} {\bibinfo  {journal}
  {Proceedings of the National Academy of Sciences}\ }\textbf {\bibinfo
  {volume} {109}},\ \bibinfo {pages} {754} (\bibinfo {year}
  {2012})}\BibitemShut {NoStop}%
\bibitem [{\citenamefont {Lemieux}\ \emph {et~al.}(2020)\citenamefont
  {Lemieux}, \citenamefont {Heim}, \citenamefont {Poulin}, \citenamefont
  {Svore},\ and\ \citenamefont {Troyer}}]{Lemieux_2020}%
  \BibitemOpen
  \bibfield  {author} {\bibinfo {author} {\bibfnamefont {J.}~\bibnamefont
  {Lemieux}}, \bibinfo {author} {\bibfnamefont {B.}~\bibnamefont {Heim}},
  \bibinfo {author} {\bibfnamefont {D.}~\bibnamefont {Poulin}}, \bibinfo
  {author} {\bibfnamefont {K.}~\bibnamefont {Svore}}, \ and\ \bibinfo {author}
  {\bibfnamefont {M.}~\bibnamefont {Troyer}},\ }\href {\doibase
  10.22331/q-2020-06-29-287} {\bibfield  {journal} {\bibinfo  {journal}
  {Quantum}\ }\textbf {\bibinfo {volume} {4}},\ \bibinfo {pages} {287}
  (\bibinfo {year} {2020})}\BibitemShut {NoStop}%
\bibitem [{\citenamefont {Montanaro}(2015)}]{Montanaro_2015}%
  \BibitemOpen
  \bibfield  {author} {\bibinfo {author} {\bibfnamefont {A.}~\bibnamefont
  {Montanaro}},\ }\href {\doibase 10.1098/rspa.2015.0301} {\bibfield  {journal}
  {\bibinfo  {journal} {Proceedings of the Royal Society A: Mathematical,
  Physical and Engineering Sciences}\ }\textbf {\bibinfo {volume} {471}},\
  \bibinfo {pages} {20150301} (\bibinfo {year} {2015})}\BibitemShut {NoStop}%
\bibitem [{\citenamefont {Harrow}\ and\ \citenamefont
  {Wei}(2020)}]{harrow2020adaptive}%
  \BibitemOpen
  \bibfield  {author} {\bibinfo {author} {\bibfnamefont {A.~W.}\ \bibnamefont
  {Harrow}}\ and\ \bibinfo {author} {\bibfnamefont {A.~Y.}\ \bibnamefont
  {Wei}},\ }in\ \href@noop {} {\emph {\bibinfo {booktitle} {Proceedings of the
  Fourteenth Annual ACM-SIAM Symposium on Discrete Algorithms}}}\ (\bibinfo
  {organization} {SIAM},\ \bibinfo {year} {2020})\ pp.\ \bibinfo {pages}
  {193--212}\BibitemShut {NoStop}%
\bibitem [{\citenamefont {Arunachalam}\ \emph {et~al.}(2022)\citenamefont
  {Arunachalam}, \citenamefont {Havlicek}, \citenamefont {Nannicini},
  \citenamefont {Temme},\ and\ \citenamefont {Wocjan}}]{Arunachalam_2022}%
  \BibitemOpen
  \bibfield  {author} {\bibinfo {author} {\bibfnamefont {S.}~\bibnamefont
  {Arunachalam}}, \bibinfo {author} {\bibfnamefont {V.}~\bibnamefont
  {Havlicek}}, \bibinfo {author} {\bibfnamefont {G.}~\bibnamefont {Nannicini}},
  \bibinfo {author} {\bibfnamefont {K.}~\bibnamefont {Temme}}, \ and\ \bibinfo
  {author} {\bibfnamefont {P.}~\bibnamefont {Wocjan}},\ }\href {\doibase
  10.22331/q-2022-09-01-789} {\bibfield  {journal} {\bibinfo  {journal}
  {Quantum}\ }\textbf {\bibinfo {volume} {6}},\ \bibinfo {pages} {789}
  (\bibinfo {year} {2022})}\BibitemShut {NoStop}%
\bibitem [{\citenamefont {Steudtner}\ \emph {et~al.}(2023)\citenamefont
  {Steudtner}, \citenamefont {Morley-Short}, \citenamefont {Pol}, \citenamefont
  {Sim}, \citenamefont {Cortes}, \citenamefont {Loipersberger}, \citenamefont
  {Parrish}, \citenamefont {Degroote}, \citenamefont {Moll}, \citenamefont
  {Santagati},\ and\ \citenamefont {Streif}}]{steudtner2023fault}%
  \BibitemOpen
  \bibfield  {author} {\bibinfo {author} {\bibfnamefont {M.}~\bibnamefont
  {Steudtner}}, \bibinfo {author} {\bibfnamefont {S.}~\bibnamefont
  {Morley-Short}}, \bibinfo {author} {\bibfnamefont {W.}~\bibnamefont {Pol}},
  \bibinfo {author} {\bibfnamefont {S.}~\bibnamefont {Sim}}, \bibinfo {author}
  {\bibfnamefont {C.~L.}\ \bibnamefont {Cortes}}, \bibinfo {author}
  {\bibfnamefont {M.}~\bibnamefont {Loipersberger}}, \bibinfo {author}
  {\bibfnamefont {R.~M.}\ \bibnamefont {Parrish}}, \bibinfo {author}
  {\bibfnamefont {M.}~\bibnamefont {Degroote}}, \bibinfo {author}
  {\bibfnamefont {N.}~\bibnamefont {Moll}}, \bibinfo {author} {\bibfnamefont
  {R.}~\bibnamefont {Santagati}}, \ and\ \bibinfo {author} {\bibfnamefont
  {M.}~\bibnamefont {Streif}},\ }\href {\doibase 10.48550/arxiv.2303.14118}
  {\bibfield  {journal} {\bibinfo  {journal} {arXiv}\ } (\bibinfo {year}
  {2023}),\ 10.48550/arxiv.2303.14118},\ \Eprint
  {http://arxiv.org/abs/2303.14118} {2303.14118} \BibitemShut {NoStop}%
\bibitem [{\citenamefont {Luis}\ and\ \citenamefont
  {Perina}(1996)}]{luis1996optimum}%
  \BibitemOpen
  \bibfield  {author} {\bibinfo {author} {\bibfnamefont {A.}~\bibnamefont
  {Luis}}\ and\ \bibinfo {author} {\bibfnamefont {J.}~\bibnamefont {Perina}},\
  }\href {\doibase 10.1103/physreva.54.4564} {\bibfield  {journal} {\bibinfo
  {journal} {Physical Review A}\ }\textbf {\bibinfo {volume} {54}},\ \bibinfo
  {pages} {4564} (\bibinfo {year} {1996})}\BibitemShut {NoStop}%
\bibitem [{\citenamefont {Dam}\ \emph {et~al.}(2007)\citenamefont {Dam},
  \citenamefont {D'Ariano}, \citenamefont {Ekert}, \citenamefont
  {Macchiavello},\ and\ \citenamefont {Mosca}}]{dam2007optimal}%
  \BibitemOpen
  \bibfield  {author} {\bibinfo {author} {\bibfnamefont {W.~v.}\ \bibnamefont
  {Dam}}, \bibinfo {author} {\bibfnamefont {G.~M.}\ \bibnamefont {D'Ariano}},
  \bibinfo {author} {\bibfnamefont {A.}~\bibnamefont {Ekert}}, \bibinfo
  {author} {\bibfnamefont {C.}~\bibnamefont {Macchiavello}}, \ and\ \bibinfo
  {author} {\bibfnamefont {M.}~\bibnamefont {Mosca}},\ }\href {\doibase
  10.1103/physrevlett.98.090501} {\bibfield  {journal} {\bibinfo  {journal}
  {Physical Review Letters}\ }\textbf {\bibinfo {volume} {98}},\ \bibinfo
  {pages} {090501} (\bibinfo {year} {2007})},\ \Eprint
  {http://arxiv.org/abs/quant-ph/0609160} {quant-ph/0609160} \BibitemShut
  {NoStop}%
\bibitem [{\citenamefont {Najafi}\ \emph {et~al.}(2023)\citenamefont {Najafi},
  \citenamefont {Costa},\ and\ \citenamefont {Berry}}]{najafi2023optimum}%
  \BibitemOpen
  \bibfield  {author} {\bibinfo {author} {\bibfnamefont {P.}~\bibnamefont
  {Najafi}}, \bibinfo {author} {\bibfnamefont {P.~C.~S.}\ \bibnamefont
  {Costa}}, \ and\ \bibinfo {author} {\bibfnamefont {D.~W.}\ \bibnamefont
  {Berry}},\ }\href {\doibase 10.48550/arxiv.2303.12503} {\bibfield  {journal}
  {\bibinfo  {journal} {arXiv}\ } (\bibinfo {year} {2023}),\
  10.48550/arxiv.2303.12503},\ \Eprint {http://arxiv.org/abs/2303.12503}
  {2303.12503} \BibitemShut {NoStop}%
\bibitem [{\citenamefont {Cleve}\ \emph {et~al.}(1998)\citenamefont {Cleve},
  \citenamefont {Ekert}, \citenamefont {Macchiavello},\ and\ \citenamefont
  {Mosca}}]{cleve1998quantum}%
  \BibitemOpen
  \bibfield  {author} {\bibinfo {author} {\bibfnamefont {R.}~\bibnamefont
  {Cleve}}, \bibinfo {author} {\bibfnamefont {A.}~\bibnamefont {Ekert}},
  \bibinfo {author} {\bibfnamefont {C.}~\bibnamefont {Macchiavello}}, \ and\
  \bibinfo {author} {\bibfnamefont {M.}~\bibnamefont {Mosca}},\ }\href
  {\doibase 10.1098/rspa.1998.0164} {\bibfield  {journal} {\bibinfo  {journal}
  {Proceedings of the Royal Society of London. Series A: Mathematical, Physical
  and Engineering Sciences}\ }\textbf {\bibinfo {volume} {454}},\ \bibinfo
  {pages} {339} (\bibinfo {year} {1998})},\ \Eprint
  {http://arxiv.org/abs/quant-ph/9708016} {quant-ph/9708016} \BibitemShut
  {NoStop}%
\bibitem [{\citenamefont {Nielsen}\ and\ \citenamefont
  {Chuang}(2010)}]{nielsen_chuang_2010}%
  \BibitemOpen
  \bibfield  {author} {\bibinfo {author} {\bibfnamefont {M.~A.}\ \bibnamefont
  {Nielsen}}\ and\ \bibinfo {author} {\bibfnamefont {I.~L.}\ \bibnamefont
  {Chuang}},\ }\href {\doibase 10.1017/CBO9780511976667} {\emph {\bibinfo
  {title} {Quantum Computation and Quantum Information: 10th Anniversary
  Edition}}}\ (\bibinfo  {publisher} {Cambridge University Press},\ \bibinfo
  {year} {2010})\BibitemShut {NoStop}%
\bibitem [{\citenamefont {Berry}\ \emph {et~al.}(2022)\citenamefont {Berry},
  \citenamefont {Su}, \citenamefont {Gyurik}, \citenamefont {King},
  \citenamefont {Basso}, \citenamefont {Barba}, \citenamefont {Rajput},
  \citenamefont {Wiebe}, \citenamefont {Dunjko},\ and\ \citenamefont
  {Babbush}}]{berry2022quantifying}%
  \BibitemOpen
  \bibfield  {author} {\bibinfo {author} {\bibfnamefont {D.~W.}\ \bibnamefont
  {Berry}}, \bibinfo {author} {\bibfnamefont {Y.}~\bibnamefont {Su}}, \bibinfo
  {author} {\bibfnamefont {C.}~\bibnamefont {Gyurik}}, \bibinfo {author}
  {\bibfnamefont {R.}~\bibnamefont {King}}, \bibinfo {author} {\bibfnamefont
  {J.}~\bibnamefont {Basso}}, \bibinfo {author} {\bibfnamefont {A.~D.~T.}\
  \bibnamefont {Barba}}, \bibinfo {author} {\bibfnamefont {A.}~\bibnamefont
  {Rajput}}, \bibinfo {author} {\bibfnamefont {N.}~\bibnamefont {Wiebe}},
  \bibinfo {author} {\bibfnamefont {V.}~\bibnamefont {Dunjko}}, \ and\ \bibinfo
  {author} {\bibfnamefont {R.}~\bibnamefont {Babbush}},\ }\href {\doibase
  10.48550/arxiv.2209.13581} {\bibfield  {journal} {\bibinfo  {journal}
  {arXiv}\ } (\bibinfo {year} {2022}),\ 10.48550/arxiv.2209.13581},\ \Eprint
  {http://arxiv.org/abs/2209.13581} {2209.13581} \BibitemShut {NoStop}%
\bibitem [{\citenamefont {Harris}(1978)}]{harris1978use}%
  \BibitemOpen
  \bibfield  {author} {\bibinfo {author} {\bibfnamefont {F.}~\bibnamefont
  {Harris}},\ }\href {\doibase 10.1109/proc.1978.10837} {\bibfield  {journal}
  {\bibinfo  {journal} {Proceedings of the IEEE}\ }\textbf {\bibinfo {volume}
  {66}},\ \bibinfo {pages} {51} (\bibinfo {year} {1978})}\BibitemShut {NoStop}%
\bibitem [{\citenamefont {Dong}\ \emph {et~al.}(2021)\citenamefont {Dong},
  \citenamefont {Meng}, \citenamefont {Whaley},\ and\ \citenamefont
  {Lin}}]{dong2021efficient}%
  \BibitemOpen
  \bibfield  {author} {\bibinfo {author} {\bibfnamefont {Y.}~\bibnamefont
  {Dong}}, \bibinfo {author} {\bibfnamefont {X.}~\bibnamefont {Meng}}, \bibinfo
  {author} {\bibfnamefont {K.~B.}\ \bibnamefont {Whaley}}, \ and\ \bibinfo
  {author} {\bibfnamefont {L.}~\bibnamefont {Lin}},\ }\href {\doibase
  10.1103/PhysRevA.103.042419} {\bibfield  {journal} {\bibinfo  {journal}
  {Phys. Rev. A}\ }\textbf {\bibinfo {volume} {103}},\ \bibinfo {pages}
  {042419} (\bibinfo {year} {2021})}\BibitemShut {NoStop}%
\bibitem [{\citenamefont {Haah}(2019)}]{haah2019product}%
  \BibitemOpen
  \bibfield  {author} {\bibinfo {author} {\bibfnamefont {J.}~\bibnamefont
  {Haah}},\ }\href {\doibase 10.22331/q-2019-10-07-190} {\bibfield  {journal}
  {\bibinfo  {journal} {Quantum}\ }\textbf {\bibinfo {volume} {3}},\ \bibinfo
  {pages} {190} (\bibinfo {year} {2019})},\ \Eprint
  {http://arxiv.org/abs/1806.10236} {1806.10236} \BibitemShut {NoStop}%
\bibitem [{\citenamefont {Chao}\ \emph {et~al.}(2020)\citenamefont {Chao},
  \citenamefont {Ding}, \citenamefont {Gilyen}, \citenamefont {Huang},\ and\
  \citenamefont {Szegedy}}]{chao2020finding}%
  \BibitemOpen
  \bibfield  {author} {\bibinfo {author} {\bibfnamefont {R.}~\bibnamefont
  {Chao}}, \bibinfo {author} {\bibfnamefont {D.}~\bibnamefont {Ding}}, \bibinfo
  {author} {\bibfnamefont {A.}~\bibnamefont {Gilyen}}, \bibinfo {author}
  {\bibfnamefont {C.}~\bibnamefont {Huang}}, \ and\ \bibinfo {author}
  {\bibfnamefont {M.}~\bibnamefont {Szegedy}},\ }\href {\doibase
  10.48550/arxiv.2003.02831} {\bibfield  {journal} {\bibinfo  {journal}
  {arXiv}\ } (\bibinfo {year} {2020}),\ 10.48550/arxiv.2003.02831},\ \Eprint
  {http://arxiv.org/abs/2003.02831} {2003.02831} \BibitemShut {NoStop}%
\bibitem [{\citenamefont {Ying}(2022)}]{ying2022stable}%
  \BibitemOpen
  \bibfield  {author} {\bibinfo {author} {\bibfnamefont {L.}~\bibnamefont
  {Ying}},\ }\href {\doibase 10.22331/q-2022-10-20-842} {\bibfield  {journal}
  {\bibinfo  {journal} {Quantum}\ }\textbf {\bibinfo {volume} {6}},\ \bibinfo
  {pages} {842} (\bibinfo {year} {2022})},\ \Eprint
  {http://arxiv.org/abs/2202.02671} {2202.02671} \BibitemShut {NoStop}%
\bibitem [{\citenamefont {Dong}\ \emph {et~al.}(2013)\citenamefont {Dong},
  \citenamefont {Wang}, \citenamefont {Meng}, \citenamefont {Ni},\ and\
  \citenamefont {Lin}}]{qsppack}%
  \BibitemOpen
  \bibfield  {author} {\bibinfo {author} {\bibfnamefont {Y.}~\bibnamefont
  {Dong}}, \bibinfo {author} {\bibfnamefont {J.}~\bibnamefont {Wang}}, \bibinfo
  {author} {\bibfnamefont {X.}~\bibnamefont {Meng}}, \bibinfo {author}
  {\bibfnamefont {H.}~\bibnamefont {Ni}}, \ and\ \bibinfo {author}
  {\bibfnamefont {L.}~\bibnamefont {Lin}},\ }\href@noop {} {\enquote {\bibinfo
  {title} {{QSP} {P}hase {F}actor {S}olvers},}\ }\bibinfo {howpublished}
  {\url{https://github.com/qsppack/QSPPACK}} (\bibinfo {year}
  {2013})\BibitemShut {NoStop}%
\bibitem [{\citenamefont {Wang}\ \emph {et~al.}(2023)\citenamefont {Wang},
  \citenamefont {Zhang}, \citenamefont {Yu},\ and\ \citenamefont
  {Wang}}]{wang2023quantum}%
  \BibitemOpen
  \bibfield  {author} {\bibinfo {author} {\bibfnamefont {Y.}~\bibnamefont
  {Wang}}, \bibinfo {author} {\bibfnamefont {L.}~\bibnamefont {Zhang}},
  \bibinfo {author} {\bibfnamefont {Z.}~\bibnamefont {Yu}}, \ and\ \bibinfo
  {author} {\bibfnamefont {X.}~\bibnamefont {Wang}},\ }\href {\doibase
  10.1103/physreva.108.062413} {\bibfield  {journal} {\bibinfo  {journal}
  {Physical Review A}\ }\textbf {\bibinfo {volume} {108}},\ \bibinfo {pages}
  {062413} (\bibinfo {year} {2023})},\ \Eprint
  {http://arxiv.org/abs/2209.14278} {2209.14278} \BibitemShut {NoStop}%
\bibitem [{\citenamefont {Motlagh}\ and\ \citenamefont
  {Wiebe}(2023)}]{motlagh2023generalized}%
  \BibitemOpen
  \bibfield  {author} {\bibinfo {author} {\bibfnamefont {D.}~\bibnamefont
  {Motlagh}}\ and\ \bibinfo {author} {\bibfnamefont {N.}~\bibnamefont
  {Wiebe}},\ }\href@noop {} {\enquote {\bibinfo {title} {Generalized quantum
  signal processing},}\ } (\bibinfo {year} {2023}),\ \Eprint
  {http://arxiv.org/abs/2308.01501} {arXiv:2308.01501 [quant-ph]} \BibitemShut
  {NoStop}%
\bibitem [{\citenamefont {O'Malley}\ \emph {et~al.}(2016)\citenamefont
  {O'Malley}, \citenamefont {Babbush}, \citenamefont {Kivlichan}, \citenamefont
  {Romero}, \citenamefont {McClean}, \citenamefont {Barends}, \citenamefont
  {Kelly}, \citenamefont {Roushan}, \citenamefont {Tranter}, \citenamefont
  {Ding}, \citenamefont {Campbell}, \citenamefont {Chen}, \citenamefont {Chen},
  \citenamefont {Chiaro}, \citenamefont {Dunsworth}, \citenamefont {Fowler},
  \citenamefont {Jeffrey}, \citenamefont {Lucero}, \citenamefont {Megrant},
  \citenamefont {Mutus}, \citenamefont {Neeley}, \citenamefont {Neill},
  \citenamefont {Quintana}, \citenamefont {Sank}, \citenamefont {Vainsencher},
  \citenamefont {Wenner}, \citenamefont {White}, \citenamefont {Coveney},
  \citenamefont {Love}, \citenamefont {Neven}, \citenamefont {Aspuru-Guzik},\
  and\ \citenamefont {Martinis}}]{omalley2016scalable}%
  \BibitemOpen
  \bibfield  {author} {\bibinfo {author} {\bibfnamefont {P.~J.~J.}\
  \bibnamefont {O'Malley}}, \bibinfo {author} {\bibfnamefont {R.}~\bibnamefont
  {Babbush}}, \bibinfo {author} {\bibfnamefont {I.~D.}\ \bibnamefont
  {Kivlichan}}, \bibinfo {author} {\bibfnamefont {J.}~\bibnamefont {Romero}},
  \bibinfo {author} {\bibfnamefont {J.~R.}\ \bibnamefont {McClean}}, \bibinfo
  {author} {\bibfnamefont {R.}~\bibnamefont {Barends}}, \bibinfo {author}
  {\bibfnamefont {J.}~\bibnamefont {Kelly}}, \bibinfo {author} {\bibfnamefont
  {P.}~\bibnamefont {Roushan}}, \bibinfo {author} {\bibfnamefont
  {A.}~\bibnamefont {Tranter}}, \bibinfo {author} {\bibfnamefont
  {N.}~\bibnamefont {Ding}}, \bibinfo {author} {\bibfnamefont {B.}~\bibnamefont
  {Campbell}}, \bibinfo {author} {\bibfnamefont {Y.}~\bibnamefont {Chen}},
  \bibinfo {author} {\bibfnamefont {Z.}~\bibnamefont {Chen}}, \bibinfo {author}
  {\bibfnamefont {B.}~\bibnamefont {Chiaro}}, \bibinfo {author} {\bibfnamefont
  {A.}~\bibnamefont {Dunsworth}}, \bibinfo {author} {\bibfnamefont {A.~G.}\
  \bibnamefont {Fowler}}, \bibinfo {author} {\bibfnamefont {E.}~\bibnamefont
  {Jeffrey}}, \bibinfo {author} {\bibfnamefont {E.}~\bibnamefont {Lucero}},
  \bibinfo {author} {\bibfnamefont {A.}~\bibnamefont {Megrant}}, \bibinfo
  {author} {\bibfnamefont {J.~Y.}\ \bibnamefont {Mutus}}, \bibinfo {author}
  {\bibfnamefont {M.}~\bibnamefont {Neeley}}, \bibinfo {author} {\bibfnamefont
  {C.}~\bibnamefont {Neill}}, \bibinfo {author} {\bibfnamefont
  {C.}~\bibnamefont {Quintana}}, \bibinfo {author} {\bibfnamefont
  {D.}~\bibnamefont {Sank}}, \bibinfo {author} {\bibfnamefont {A.}~\bibnamefont
  {Vainsencher}}, \bibinfo {author} {\bibfnamefont {J.}~\bibnamefont {Wenner}},
  \bibinfo {author} {\bibfnamefont {T.~C.}\ \bibnamefont {White}}, \bibinfo
  {author} {\bibfnamefont {P.~V.}\ \bibnamefont {Coveney}}, \bibinfo {author}
  {\bibfnamefont {P.~J.}\ \bibnamefont {Love}}, \bibinfo {author}
  {\bibfnamefont {H.}~\bibnamefont {Neven}}, \bibinfo {author} {\bibfnamefont
  {A.}~\bibnamefont {Aspuru-Guzik}}, \ and\ \bibinfo {author} {\bibfnamefont
  {J.~M.}\ \bibnamefont {Martinis}},\ }\href {\doibase
  10.1103/PhysRevX.6.031007} {\bibfield  {journal} {\bibinfo  {journal} {Phys.
  Rev. X}\ }\textbf {\bibinfo {volume} {6}},\ \bibinfo {pages} {031007}
  (\bibinfo {year} {2016})}\BibitemShut {NoStop}%
\bibitem [{\citenamefont {Apel}\ \emph {et~al.}(2024)\citenamefont {Apel},
  \citenamefont {Lemieux},\ and\ \citenamefont {Steudner}}]{apel2024kaiser}%
  \BibitemOpen
  \bibfield  {author} {\bibinfo {author} {\bibfnamefont {H.}~\bibnamefont
  {Apel}}, \bibinfo {author} {\bibfnamefont {J.}~\bibnamefont {Lemieux}}, \
  and\ \bibinfo {author} {\bibfnamefont {M.}~\bibnamefont {Steudner}},\
  }\href@noop {} {}\bibinfo {howpublished} {In Preparation} (\bibinfo {year}
  {2024})\BibitemShut {NoStop}%
\bibitem [{\citenamefont {Low}\ and\ \citenamefont
  {Chuang}(2019)}]{Low2019hamiltonian}%
  \BibitemOpen
  \bibfield  {author} {\bibinfo {author} {\bibfnamefont {G.~H.}\ \bibnamefont
  {Low}}\ and\ \bibinfo {author} {\bibfnamefont {I.~L.}\ \bibnamefont
  {Chuang}},\ }\href {\doibase 10.22331/q-2019-07-12-163} {\bibfield  {journal}
  {\bibinfo  {journal} {{Quantum}}\ }\textbf {\bibinfo {volume} {3}},\ \bibinfo
  {pages} {163} (\bibinfo {year} {2019})}\BibitemShut {NoStop}%
\bibitem [{\citenamefont {Low}\ and\ \citenamefont
  {Chuang}(2017)}]{low2017optimal}%
  \BibitemOpen
  \bibfield  {author} {\bibinfo {author} {\bibfnamefont {G.~H.}\ \bibnamefont
  {Low}}\ and\ \bibinfo {author} {\bibfnamefont {I.~L.}\ \bibnamefont
  {Chuang}},\ }\href {\doibase 10.1103/PhysRevLett.118.010501} {\bibfield
  {journal} {\bibinfo  {journal} {Phys. Rev. Lett.}\ }\textbf {\bibinfo
  {volume} {118}},\ \bibinfo {pages} {010501} (\bibinfo {year}
  {2017})}\BibitemShut {NoStop}%
\bibitem [{\citenamefont {McArdle}\ \emph {et~al.}(2022)\citenamefont
  {McArdle}, \citenamefont {Gilyen},\ and\ \citenamefont
  {Berta}}]{mcardle2022quantum}%
  \BibitemOpen
  \bibfield  {author} {\bibinfo {author} {\bibfnamefont {S.}~\bibnamefont
  {McArdle}}, \bibinfo {author} {\bibfnamefont {A.}~\bibnamefont {Gilyen}}, \
  and\ \bibinfo {author} {\bibfnamefont {M.}~\bibnamefont {Berta}},\ }\href
  {\doibase 10.48550/arxiv.2210.14892} {\bibfield  {journal} {\bibinfo
  {journal} {arXiv}\ } (\bibinfo {year} {2022}),\ 10.48550/arxiv.2210.14892},\
  \Eprint {http://arxiv.org/abs/2210.14892} {2210.14892} \BibitemShut {NoStop}%
\bibitem [{\citenamefont {Low}\ \emph {et~al.}(2018)\citenamefont {Low},
  \citenamefont {Kliuchnikov},\ and\ \citenamefont
  {Schaeffer}}]{low2018trading}%
  \BibitemOpen
  \bibfield  {author} {\bibinfo {author} {\bibfnamefont {G.~H.}\ \bibnamefont
  {Low}}, \bibinfo {author} {\bibfnamefont {V.}~\bibnamefont {Kliuchnikov}}, \
  and\ \bibinfo {author} {\bibfnamefont {L.}~\bibnamefont {Schaeffer}},\ }\href
  {\doibase 10.48550/arxiv.1812.00954} {\bibfield  {journal} {\bibinfo
  {journal} {arXiv}\ } (\bibinfo {year} {2018}),\ 10.48550/arxiv.1812.00954},\
  \Eprint {http://arxiv.org/abs/1812.00954} {1812.00954} \BibitemShut {NoStop}%
\bibitem [{\citenamefont {Sanders}\ \emph {et~al.}(2020)\citenamefont
  {Sanders}, \citenamefont {Berry}, \citenamefont {Costa}, \citenamefont
  {Tessler}, \citenamefont {Wiebe}, \citenamefont {Gidney}, \citenamefont
  {Neven},\ and\ \citenamefont {Babbush}}]{sanders2020compilation}%
  \BibitemOpen
  \bibfield  {author} {\bibinfo {author} {\bibfnamefont {Y.~R.}\ \bibnamefont
  {Sanders}}, \bibinfo {author} {\bibfnamefont {D.~W.}\ \bibnamefont {Berry}},
  \bibinfo {author} {\bibfnamefont {P.~C.~S.}\ \bibnamefont {Costa}}, \bibinfo
  {author} {\bibfnamefont {L.~W.}\ \bibnamefont {Tessler}}, \bibinfo {author}
  {\bibfnamefont {N.}~\bibnamefont {Wiebe}}, \bibinfo {author} {\bibfnamefont
  {C.}~\bibnamefont {Gidney}}, \bibinfo {author} {\bibfnamefont
  {H.}~\bibnamefont {Neven}}, \ and\ \bibinfo {author} {\bibfnamefont
  {R.}~\bibnamefont {Babbush}},\ }\href {\doibase 10.1103/prxquantum.1.020312}
  {\bibfield  {journal} {\bibinfo  {journal} {PRX Quantum}\ }\textbf {\bibinfo
  {volume} {1}} (\bibinfo {year} {2020}),\ 10.1103/prxquantum.1.020312},\
  \Eprint {http://arxiv.org/abs/2007.07391} {2007.07391} \BibitemShut {NoStop}%
\bibitem [{\citenamefont {Gidney}(2018{\natexlab{b}})}]{gidney2018halving}%
  \BibitemOpen
  \bibfield  {author} {\bibinfo {author} {\bibfnamefont {C.}~\bibnamefont
  {Gidney}},\ }\href {\doibase 10.22331/q-2018-06-18-74} {\bibfield  {journal}
  {\bibinfo  {journal} {Quantum}\ }\textbf {\bibinfo {volume} {2}},\ \bibinfo
  {pages} {74} (\bibinfo {year} {2018}{\natexlab{b}})},\ \Eprint
  {http://arxiv.org/abs/1709.06648} {1709.06648} \BibitemShut {NoStop}%
\end{thebibliography}%


\appendix

\section{Circuit simulation details}\label{app:sim_details}

\subsection{QSVT QPE}

\begin{sidewaysfigure}[ht]
\includegraphics[width=\textwidth]{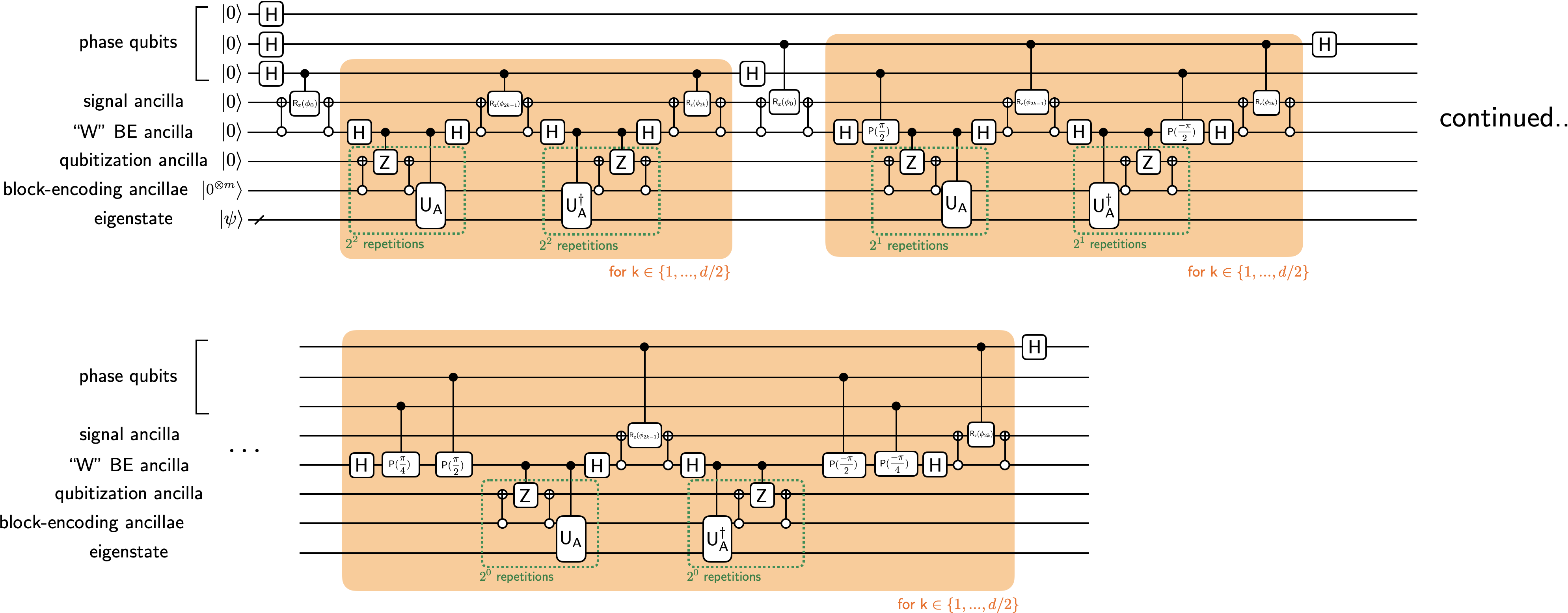}
\caption{Numerically verified QSVT QPE circuit that obtains the eigenphase of a qubitization iterate as $\exp(i \arccos(\lambda))$, with $\lambda$ the eigenvalue of interest for the block-encoded matrix $A$. $U_A$ denotes a block encoding of $A$, which is combined with a reflection about the all-zero sector of the block encoding ancilla subspace to form the qubitization iterate. This in turn is controlled on an additional ancilla sandwiched by Hadamard gates and (depending on the iteration of the QPE protocol) a set of rotation gates, with the latter used to counterrotate the phase bits measured in previous iterations of the QPE routine. This construction results in the block encoding of a new matrix $W_j(\theta) := \frac{1}{2}(I + \exp(-2\pi i \theta)(Z_\Pi U_A)^{2^j})$ using a single block encoding ancilla (labelled ``W'' BE ancilla in the circuit). Finally, the $R_z$ gates that implement the QSVT phase rotations are applied using projectors into the zero subspace of the "W" block encoding -- these rotations are controlled on the QPE phase register in order to load the estimate of the eigenphase into it. Since we are implementing QSVT, the protocol alternates between applications of $W$ and applications of $W^\dagger$. The units within the orange boxes correspond to the repetitions necessary to realize the QSVT polynomial -- each of the $d+1$ phases applied within each box (plus the single rotation before each box) could, in principle, be unique (they are obtained via optimization for the desired target polynomial), but the set of phases for each box are the same. For this specific circuit, 3 phase bits are used, but the simulations and verifications were performed with various numbers of phase qubits up to 11.}
\label{fig:qsvt_qpe_circuit}
\end{sidewaysfigure}

In Section~\ref{sec:sim_results}, we present the results of state vector simulations of the different QPE routines, including an implementation of the QSVT QPE routine formulated in Ref.~\cite{martyn2021grand}. In this implementation, we identified four small modifications to the circuit presented in Fig.~12 of Ref.~\cite{martyn2021grand} that were necessary to obtain valid results. In this section, we detail the specific circuit we used and explicitly note these modifications. The circuit used to implement the QSVT QPE routine is shown in Fig.~\ref{fig:qsvt_qpe_circuit}. The main differences from the circuit in Fig.~12 of Ref.~\cite{martyn2021grand} are:

\begin{enumerate}
    \item In our circuit we use qubitization to obtain the unitary for phase estimation.
    \item The controlled phase rotations to remove the previously measured phase bits are applied using the opposite bit ordering in our circuit compared with Fig.~12 of Ref.~\cite{martyn2021grand} -- here, the $\pi/2$ rotation is applied on the most significant bit measured, the $\pi/4$ rotation on the second most significant bit etc., while in Ref.~\cite{martyn2021grand} the controls were ordered in the opposite way.
    \item The degree of the polynomial in QSVT is (at most) equal to the sum of the number of applications of $U_A$ and $U_A^\dagger$, so to realize a $d-$degree polynomial, we only need to repeat the circuit in the orange box $d/2$ times rather than $d$ times in Fig.~12 of Ref.~\cite{martyn2021grand}.
    \item The polynomial we use (Eq.~\eqref{eq:poly_approx}) differs by a sign from that given in Eq.~(69) of Ref.~\cite{martyn2021grand}.
\end{enumerate}

The circuit was numerically verified using small QPE instances including the two-qubit Hamiltonian for molecular hydrogen over various bond lengths considered in \cite{omalley2016scalable}.

\subsection{QPE with window functions}\label{app:sim_details_windows}

\begin{figure}[h]
\centering
\includegraphics[width=0.8\textwidth]{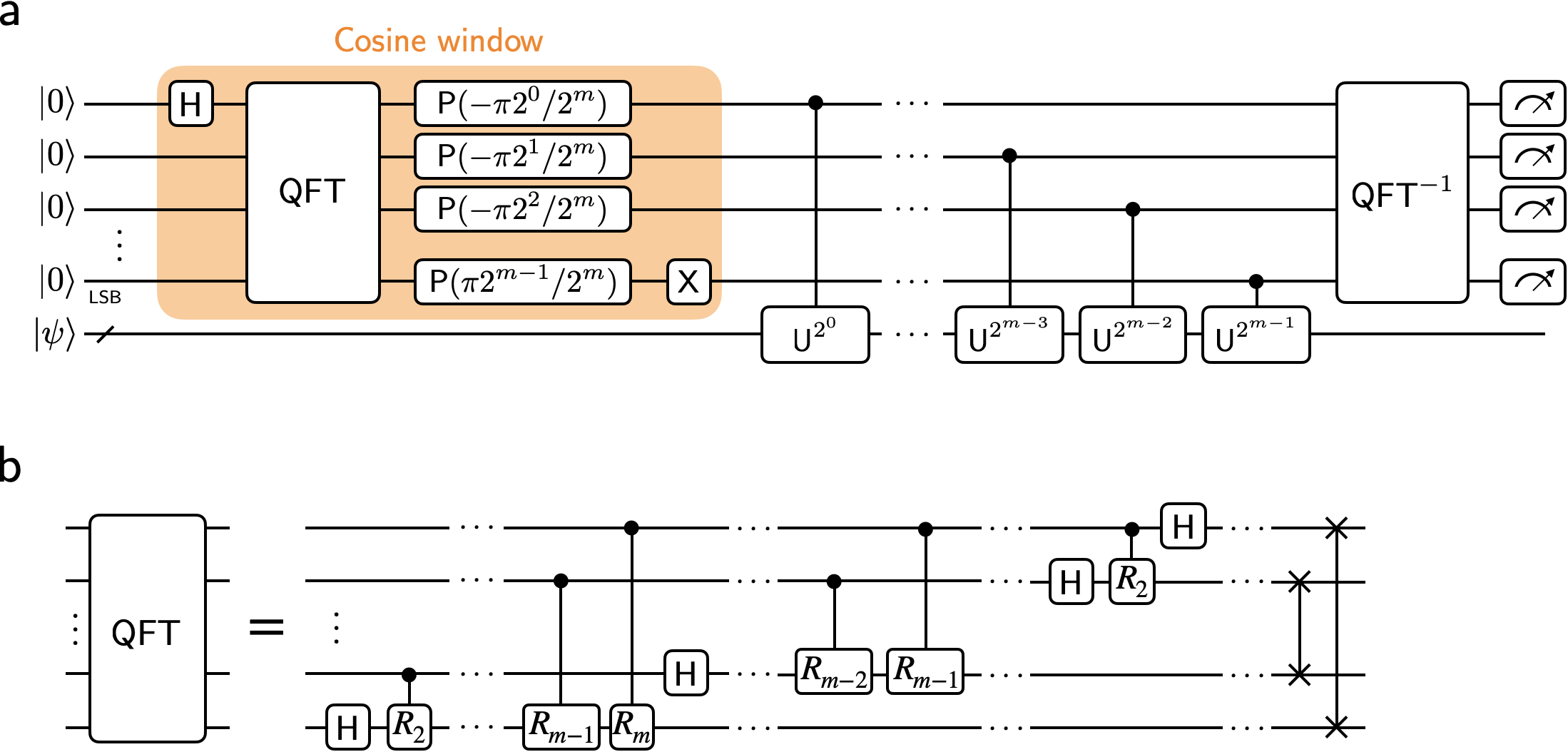}
\caption{(a) Circuit for QPE using cosine window \cite{rendon2022effects}. The lowermost qubit in the phase register is the least significant bit (LSB). Gate $\textsf{P}$ is the phase gate, defined in Eq.~\ref{eq:Pgate}.
(b) Circuit for QFT. Note that in Ref.~\cite{rendon2022effects}, this operation is denoted as QFT$^{-1}$.}
\label{fig:cosine_qpe_circuit}
\end{figure}

For simulating QPE using a window function, we show the circuit for preparing the cosine window state \cite{rendon2022effects} in Figure~\ref{fig:cosine_qpe_circuit}. As the cosine window has an exact construction and is not parametrized, no further considerations were needed to implement the simulations. The rectangular window function was even simpler to implement, consisting only of Hadamard gates on the phase register.

In order to simulate QPE using the Kaiser window, there are two considerations that must be taken into account. Firstly, a choice of the parameter $\alpha$ must be made in order to obtain a concrete definition for a Kaiser window to prepare. Secondly, we have to decide on the specific implementation details for the state preparation.
\begingroup
\begin{table}[]
\setlength{\tabcolsep}{6pt} 
\renewcommand{\arraystretch}{1.3} 
\centering
\begin{tabular}{|c|c|}
\hline
\textbf{Number of additional phase qubits} & Best $\alpha$ found \\ \hline
0 & 0  \\ 
1 & 6  \\ 
2 & 13  \\ 
3 & 25 \\ 
4 & 51  \\ 
5 & $100^*$  \\ 
6 & $100^*$  \\ \hline
\end{tabular}
\caption{Table of Kaiser window $\alpha$ values obtained through 
optimization of the QPE success probability \cite{apel2024kaiser}. 
The asterisks next to the values for 5 and 6 additional phase qubits indicate that these values are only approximate guesses rather than fully optimized values -- the optimization for these simulations was unstable since the success probabilities were high enough to be limited by floating point precision errors. 
}
\label{tab:best_alphas}
\end{table}
\endgroup
For the choice of $\alpha$, we find that the more additional phase qubits are added and then thrown away in the QPE routine, the higher the value of $\alpha$ that can be chosen. Additionally, increasing $\alpha$ leads to higher success probabilities up to a certain threshold, after which the success probabilities fall off dramatically. 
From a separate analysis and optimization of the Kaiser window \cite{apel2024kaiser}, we report best found values of $\alpha$ for each additional number of phase qubits but make no claim about the optimality of these values. 
We verified these numbers using a numerical optimization of $\alpha$ and obtained qualitatively similar results.
It is likely that different values of $\alpha$ are optimal for different applications, and the investigation of this could be a worthwhile route for future work.

For implementing the Kaiser window in our state vector simulation, we simply injected the (normalized) Kaiser window state into the phase register of the QPE simulation. While this ignores the cost of the window state, it correctly simulates the impact of applying the window state to the QPE algorithm. We leave the concrete cost analysis of the Kaiser window state to future work.

\section{Additional phase qubits in QPE using the Kaiser window state}\label{app:additional_qubits_kaiser}

In this section, we sketch out the proof for estimating the number of additional phase qubits 
needed to obtain accurate eigenphases in QPE using
Kaiser windows by extending the procedure in \cite{cleve1998quantum}.
Our sketch is based on Appendix C of \cite{berry2022quantifying}, and we strongly recommend 
that 
readers 
review this reference.

In the standard procedure from \cite{cleve1998quantum}, to derive the number of additional qubits, one first considers the expression for the amplitude of 
the 
state $\ket{(a-t) \mod 2^n}$, where $a$ is the closest $n$-bit integer approximation to some phase $\phi$, and $-2^{n-1} \leq t < 2^{n-1}$. Here, we treat $t$ as a dummy index used to compute the deviation from the nearest integer $a$. Eventually, $t$ will be substituted to compute the total probability of being at least some integer $k$ far away from $a$. We refer to the amplitude of state $\ket{(a-t) \mod 2^n}$ as $\alpha_t$ (not to be confused with the Kaiser window parameter $\alpha$). For simpler windows like the rectangular window, the procedure for computing the number of additional phase bits continues by simplifying and bounding $\alpha_t$ before summing the squared amplitudes corresponding to states that are within some distance $k/2^n$ away from $a$ for integer $k$. 
This is equivalent to computing the tails of this probability distribution (in terms of $k$) and subtracting from unit probability. We set this expression equal to some target probability $1-\delta$ then solve for $k$. We now have an expression for $k$ in terms of $\delta$ but would like to express the number of additional phase qubits in terms of $\delta$.
For this, we relate the two quantities, $k$ and the number of additional phase qubits: we set the distance $k/2^n$ equal to a target precision $\frac{1}{2^{m+1}}$, where $m \leq n$.
We define $p=n-m$ as the number of additional phase qubits and solve for $p$ using the expression for $k$ in terms of $\delta$ and noting that $k=2^{n-m-1}=2^{p-1}$. 

For the Kaiser window, writing an analytical expression for $\alpha_t$ is less trivial. In \cite{berry2022quantifying}, the authors model the probability distribution by applying the Laplace approximation method and estimate the normalization constant of the window state. 
This method approximates the probability distribution using a normal distribution given that the distribution is well-behaved (i.e. symmetric and unimodal). 
Following the procedure in \cite{berry2022quantifying}, using the approximated normalization constant and noting that the first zero or sidelobe of the distribution occurs at $(\pi/(2^{n-1}))\sqrt{1+\alpha^2}$ (for the Kaiser window parameter $\alpha$), the authors integrate the probabilities over the tails of the distribution and set this total probability (of failure) to some $\delta$. This allows one to solve for the Kaiser window parameter $\alpha$:

\begin{align}\label{eq:alpha}
    \alpha = (1/2\pi) \ln (1/\delta) + \mathcal{O}(\ln \ln (1/\delta)).
\end{align}

With an expression for $\alpha$, we now set the confidence interval $(\pi/(2^{n-1}))\sqrt{1+\alpha^2}$ equal to some target precision $\frac{1}{2^{m+1}}$ as was done above for the rectangular window function, and we again solve for $p=n-m$:

\begin{align}
\text{target precision} &= \frac{\pi}{2^{n-1}} \sqrt{1+\alpha^2} \\
\frac{1}{2^{m+1}} &= \frac{\pi}{2^{n-1}} \sqrt{1+\alpha^2} \\
2^{n-1-m-1} &= \pi \sqrt{1+\alpha^2} \\
2^{p-2} &= \pi \sqrt{1+\alpha^2} \\
2^{p-2} &\approx \pi \big( (1/2\pi) \ln(1/\delta) + \mathcal{O}(\ln \ln (1/\delta)) \big) \label{eq:alpha_approx}\\
p &= \mathcal{O} (\log \log 1/\delta). \label{eq:p_kaiser}
\end{align}

In \cite{berry2022quantifying}, they note that the $\ln\ln$ term in \ref{eq:alpha} is larger than the error for approximating $\sqrt{1+\alpha^2}$ using $\alpha$, justifying the approximation in Eq.~\ref{eq:alpha_approx}. In the end, we note that $p = \mathcal{O} (\log \log 1/\delta)$.

While we did not use this derived value to choose the number of additional qubits in our numerical simulations, we numerically verified that the Kaiser window does in fact use fewer additional phase qubits than the cosine window to achieve comparable success probabilities (beyond four qubits). Alternatively, assuming the same number of additional qubits, using the Kaiser window would correspond to a higher success probability than that obtained using the cosine window, beyond some crossover point.

\section{Quantum singular value transformation (QSVT)}\label{sec:qsvt_overview}
In this appendix, a brief overview of the quantum singular value transform subroutine is given. Many quantum algorithms can be defined in terms of functions of matrices. For example, the time evolution operator may be thought of as applying the function $f(x) := \exp(i x t)$ onto a Hamiltonian $H$. A general method for generating matrix functions on a quantum computer is given by the {\it quantum singular value transform} (QSVT)~\cite{gilyen2019quantum}. QSVT uses two primitive operations: controlled rotation gates and a block encoding oracle~\cite{Low2019hamiltonian}, which takes a target matrix $A$ and realizes a unitary block encoding operator
\begin{equation}
    U_A := \begin{bmatrix}
A & *\\
* & *
\end{bmatrix} \ ,
\end{equation}
where the terms labelled $*$ are left undefined and can take on any values. The key insight underpinning the quantum singular value transform is that the block encoding operator may be written as a direct sum of projectors into a set of invariant subspaces of the form $\{|0^m\rangle|\lambda_i\rangle, |\perp_i\rangle\}$, where $m$ is the number of ancillae qubits used to encode the matrix $A$, $|\lambda_i\rangle$ is an eigenstate of $A$ and $|\perp_i\rangle$ is orthogonal to $|0^m\rangle|\nu_i\rangle$. Since the full operator $U_A$ is defined in terms of a direct sum over all eigenstates of $A$, the result of applying the block encoding oracle onto {\it any} state of the form $|0^m\rangle|\psi\rangle$ can be understood in terms of $2 \times 2$ matrices whose elements correspond to the projectors into this invariant subspace.

This insight may be combined with the notion of {\it quantum signal processing}~\cite{low2017optimal}, a technique in which single qubit rotation gates are interspersed between applications of a single qubit unitary matrix in order to transform the output of that matrix as
\begin{equation}
    f(U) = Rz(\phi_0)\prod_{j=1}^d U Rz(\phi_j) \ ,
\end{equation}
where $d$ is the number of applications of the unitary matrix which upper bounds the degree of the matrix polynomial that can be realized. By choosing an appropriate set of rotation angles $\{\phi_i\}$, different polynomial functions of the original matrix can be implemented. The technique can be extended to arbitrary matrix transformations by defining a rotation operator $Rz_{\Pi}$ whose eigenstates are the projectors into the invariant subspaces defined above. In practice, this is accomplished by using an $Rz$ gate sandwiched between \textsc{not} gates controlled on the all-zero sector of the block encoding ancillae. The resulting gate sequence is given as
\begin{equation}
    Rz_{\Pi}(\phi_0)\prod_{j=1}^{d/2} U_A Rz_{\Pi}(\phi_j) U_A^\dagger Rz_{\Pi}(\phi_{j+1}) \ .
\end{equation}
For Hermitian matrices $A$, the result of applying this sequence of operators is a block encoding of a matrix function defined over the eigenvalues of $A$,
\begin{equation}
    f(A) := \sum_{k=0}^{\operatorname{dim}(A)-1} f(\lambda_k)|\lambda_k\rangle\langle\lambda_k| \ .
\end{equation}
When $A$ is not Hermitian, the sequence is gates is altered slightly such that we alternate between applications of $U_A$ and $U_A^\dagger$, with the result being a block encoding of a function over the {\it singular values} of $A$ instead:
\begin{equation}
    f(A) := \sum_{k=0}^{\operatorname{dim}(A)-1} f(\sigma_k)|v_k\rangle\langle w_k| \ ,
\end{equation}
where $A=\sum_{k=0}^{\operatorname{dim}(A)-1} \sigma_k|v_k\rangle\langle w_k|$ is the singular value decomposition of $A$. In both cases, the protocol is successful if the block encoding ancillae are measured to be in the all-zero state. For the remainder of this work, we will focus on QSVT.

There are several constraints on the classes of polynomial functions that can be realized via QSVT, most notably:

\begin{itemize}
    \item[1.] The degree of the polynomial must be less than or equal to $d$, the number of applications of $U_A$.
    \item[2.] The parity of the polynomial must match the parity of $d$.
    \item[3.] For all $x$ such that $|x| \leq 1$, $|f(x)| \leq 1$.
\end{itemize}

There are other conventions for QSVT that can be used, but these 
differ by
basis changes from the convention chosen here.
A recent work~\cite{motlagh2023generalized} discovered that
QSP (and therefore QSVT) can be generalized to relax the first two of these constraints; since the target function considered in this work naturally accommodates these constraints, we have not included a discussion here and only note the new work for completeness.

\section{Choice of QSVT function in QSVT QPE}\label{app:qsvt_function_choice}
We outline the choice of QSVT function used in this work (taken from Ref.~\cite{martyn2021grand}).
As mentioned in the main text, the QSVT QPE routine builds upon iterative QPE, meaning that the results of previous QPE iterations are rotated out of the eigenphase being measured. We then seek some transformation of the singular values of the resulting unitary such that at each step $k$ of the QPE routine, if the $k$th bit of the ideal eigenphase (after rounding to account for the bit truncation to $m$ bits) is 0, we kick back a phase of 1 to the phase register and if the $k$th bit is 1, we kick back a phase of $-1$. Intuitively, it is clear that this function should be a sign function, but the specific choice of sign function requires a more careful construction of a particular block encoding whose singular values perform the desired kickback.

At step $k$ of the QPE protocol a unitary block encoding of the matrix $A_k(\phi)$ is implemented using a Hadamard test on the target unitary with the previously measured eigenphases rotated out:
\begin{equation}
    A_k(\phi) := \frac{1}{2}\left(\mathbb{I} - \exp(-2\pi i 0.0\varphi_{k-1}\varphi_{k-2}...)\exp(2 \pi i 2^{m-k} \phi) \right) \ ,
\end{equation}
where $\varphi_{j}$ are the $j$th-bit approximate values of $\phi$ obtained from previous iterations. This unitary $A_k(\phi)$ is a block encoding that encodes the unitary of interest $U$, for which we want to obtain an estimate of the eigenvalue $\phi$. $A_k(\phi)$ has singular values
\begin{align}
\sigma_k &= \frac{1}{2} | 1 - \exp(-2\pi i 0.0 \varphi_{k-1} \varphi_{k-2} ...) \exp(2 \pi i 2^k \phi) | \\
 &= \frac{1}{2}  | \cos( \pi 2^{m-k} \phi - 0.0 \varphi_{k-1} \varphi_{k-2} ...) | \ .
\end{align}
Let us consider the case where $\phi$ can be exactly represented using $m$ bits. In this case, at the zeroth iteration of the QPE protocol, the singular value $\sigma_0=|\cos(\pi 0.\phi_m)|$ is 1 if $\phi_m=0$ and 0 if $\phi_m=1$. The value $\phi_m$ can then be deterministically loaded into the zeroth bit of the phase register by implementing  $1-\sigma_0$ controlled on the phase qubit using a Hadamard test. 

In general, however, $\phi$ is not exactly expressible using $m$ bits and so this protocol will not be deterministic. By modifying the transforming function, the protocol can be adapted to be deterministic for any phase: note that the most ambiguous value of $\phi$ that can occur is $0.\phi_m011...1 < 0.\phi_m1$. That is, the true phase lies precisely between our desired bin and another bin. We therefore want to realize a function $f(x)$ such that 
\begin{equation}
    f(x) = \begin{cases} 0 & \mbox{if } x > \frac{1}{\sqrt{2}} \\ 1 & \mbox{if } x < \frac{1}{\sqrt{2}} \end{cases} \ ,
\end{equation}
where the inputs $x$ to the function for QSVT are the singular values of the matrix that we are seeking to transform.
This comes from the fact that the ambiguity has a maximum rounding error in the eigenphase of $1/4$ and $\cos(\pi/4) = 1\sqrt{2}$. The function that has this property is the shifted sign function
\begin{equation}
    f(x) = \Theta\left(\frac{1}{\sqrt{2}} - x\right) \ ,
\end{equation}
with $\Theta(x)$ denoting the sign function, using the notation of Ref.~\cite{martyn2021grand} (not to be confused with the Heaviside step function). We can use QSVT to obtain a polynomial approximation of this function, which will allow for the value of the $m$th bit to be loaded into the phase register. At each subsequent step, we first rotate out the phase at the bits we have previously calculated using controlled rotations before repeating the QSVT to load the next bit into the phase register. This allows us to reduce the ambiguity when we cannot represent the true eigenphase with $m$ bits, and thereby increase the success probability of QPE. 

\section{Back-of-the-envelope gate cost for preparing the window states}
\label{sec:envelope}

To fairly compare the resource costs of the QSVT QPE and the standard QPE using additional phase qubits, the cost of the initial window state preparation must be taken into account. As reported in Fig. \ref{fig:be_cost_5_bits}, the number of queries to the block encoding in QPE is far larger in the QSVT version of QPE than in the version using additional phase qubits. So if we are to determine which version has a lower complexity in practice, we effectively need to determine how expensive preparing window states is in comparison to implementing block encodings and calling them many times for some interesting representative instance size. As a representative instance size, we choose to look at the cost of computing the ground state for the molecule FeMoco, a common benchmark in the resource estimation literature \cite{babbush2018encoding, reiher2017elucidating, lee2021even} as it is widely believed to be beyond the reach of classical computation due to its strongly-correlated properties, and for which many detailed resource costings exist.

To estimate the gate complexity of these state preparations, we must assume a particular cost model. Because QPE is a large depth circuit, we assume the cost model of a fault-tolerant quantum computer, where the cost of executing non-Clifford gates far exceeds that of Clifford gates. Thus, we will determine the non-Clifford gate complexity; in particular, the Toffoli complexity.

As of the writing of this manuscript, the lowest quoted Toffoli complexity in the literature for block encoding the Hamiltonian for the FeMoco molecule is 16923 Toffolis ($\mathcal{O}(10^5)$), and the total cost of estimating a ground state energy to chemical accuracy is $\num{3.2e10}$ Toffolis \cite{lee2021even}. The rectangular window can be prepared using only Hadamards, and so, it is effectively free. The cost of the cosine window is $\tilde{\mathcal{O}}(m)$ for $m$ phase qubits \cite{babbush2018encoding}. For the FeMoco Hamiltonian, $m = 20$ \cite{lee2021even}. Assuming we are using an additional $p = 5$ qubits, the total cost for this preparation scales linearly with $25$ qubits.  This cost is negligible in comparison to the cost of the block encoding.

As far as we know, there are no quoted asymptotic or numeric complexities for the cost of preparing the Kaiser window. Thus, here we present back-of-the-envelope costs for two methods to prepare this state: arbitrary state preparation and state preparation via Quantum Signal Processing (QSP) \cite{mcardle2022quantum}. 

We can roughly estimate the Toffoli complexity for performing arbitrary state preparation, where we classically pre-compute the amplitudes given in Table \ref{tab:window_functions} and coherently load them onto the quantum computer. A variety of methods exist in the literature, the most performant of which scale as $\tilde{\mathcal{O}}(N + \log N)$ for preparing $N$ amplitudes, including a technique called ``alias sampling'' (introduced in \cite{babbush2018encoding}) and another method we refer to as ``LKS'' (for Low, Kliuchnikov, and Schaeffer) which uses a cascade of data-loaders and adders (introduced in \cite{low2018trading}). Because we do not exactly uncompute the window state preparation at the end of QPE, we cannot employ methods that entangle the prepared state with garbage, ancillary qubits. This rules out alias sampling, and so we analyze the cost of LKS state preparation. LKS requires a bits of precision parameter for how accurately to represent each prepared amplitude. We take a conservative estimate of machine precision, $32$ bits. For a QPE phase register of $25$ qubits, using $32$ bits of precision for each prepared amplitude in the Kaiser window state, LKS gives a Toffoli complexity of order $\mathcal{O}(10^6)$. 
Though it is an order of magnitude more expensive than the the block encoding of FeMoco, this state is prepared exactly once at the beginning of the QPE, and thus, is a sub-leading cost in comparison to the total $\num{3.2e10}$ Toffoli complexity from the roughly $\mathcal{O}(10^6)$ queries to the block encoding iterate.

Alternatively, one could prepare the Kaiser window state using QSP methods, as detailed in \cite{mcardle2022quantum}. There, we must make $\mathcal{O}(\alpha + \ln(\Delta^{-1}))$ many queries to the circuit shown in Fig. 1b. in \cite{mcardle2022quantum}, which in turn makes queries to the circuit shown in Fig. 1a. in the same reference, where $\alpha$ is the window parameter value for the Kaiser window, and $\Delta$ is the error in approximation. For comparison with LKS, we can take $\ln(\Delta^{-1})$ to be equal to $32$. For a conservative estimate, we take a high value of $\alpha = 100$. The circuit depicted in Fig. 1a in \cite{mcardle2022quantum} can be prepared using addition with a phase gradient state \cite{sanders2020compilation, gidney2018halving}. Addition over $n$ qubits has cost $n - 1$ \cite{gidney2018halving}, so this costs $24$ Toffolis for our instance size (the total number of phase qubits). The circuit in Fig. 1b. makes 3 queries to this circuit, for a total of $72$ Toffolis. Now, we can multiply this count by the number of total queries given above in terms of $\alpha$ and $\ln(\Delta^{-1})$, which we take to be $100$ and $32$ (respectively). This gives us a total of $\mathcal{O}(10^4)$ Toffolis, an order of magnitude fewer Toffolis than the block encoding of the FeMoco Hamiltonian.

We have not performed a detailed, explicit compilation of preparing the Kaiser window state. Nor have we spent any time looking into alternative state preparation methods as this is not the focus of this work. Even still, the methods quoted above yield counts that are low enough to be negligible in comparison to the total cost of querying block encodings of Hamiltonians for interesting systems, justifying the comparison of QSVT QPE and textbook QPE with additional phase qubits to be done only in terms of query complexity.

\section{Asymmetry of the QSVT QPE Success Probability}\label{app:qsvt_succ_prob_assymetry}
\begin{figure}[h]
\centering
\includegraphics[width=0.6\textwidth]{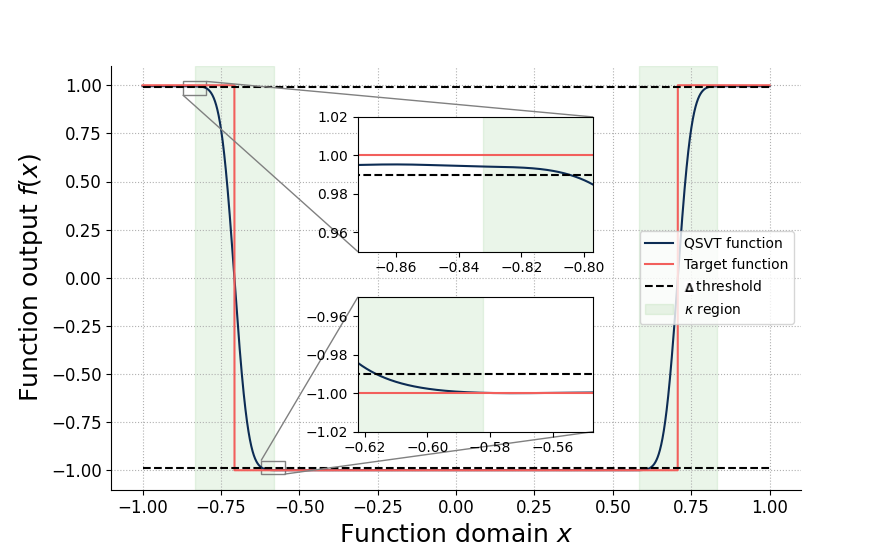}
\caption{Plot of the target shifted sign function for QSVT QPE, showing both the approximate target function Eq.~\eqref{eq:poly_approx} (red line) and the output from the function resulting from the optimized QSVT phases (dark blue line), with the target $\kappa$ region and $\Delta$ values shown as green shading and black dashed lines respectively. The zoomed insets show that the optimized function approximates the ideal function much more closely for the $-1$ branches of the target function than for the $+1$, resulting in an asymmetry in the output of QSVT QPE.}
\label{fig:QSVT_QPE_function_asymmetry}
\end{figure}

The success probability of QPE arising from bit discretization error is periodic with period $1/2^{n}$ for $n = m + p$ total bits of precision. While this is true in theory for all of the QPE routines explored here, in practice QSVT QPE is {\it not} periodic. The reason for this lies in the choice of polynomial decomposition used for the target function. The ideal shifted sign function (Eq.~\eqref{eq:poly_approx} with $\Delta = \kappa = 0$) is symmetric with respect to the different bit values $0$ and $1$, meaning that QSVT QPE using this ideal function should be periodic. However, it is not possible to use the ideal shifted sign function in a real subroutine, since that would necessitate an infinite number of terms – the {\it imperfect} shifted sign function resulting from the finite order polynomial decomposition is {\it not} symmetric with respect to these bit values. Rather, as shown in Fig.~\ref{fig:QSVT_QPE_function_asymmetry}, the deviation from the ideal values for the $1$ bits is higher than the corresponding deviation for the zero bits.

\begin{figure}[h]
\centering
\includegraphics[width=0.6\textwidth]{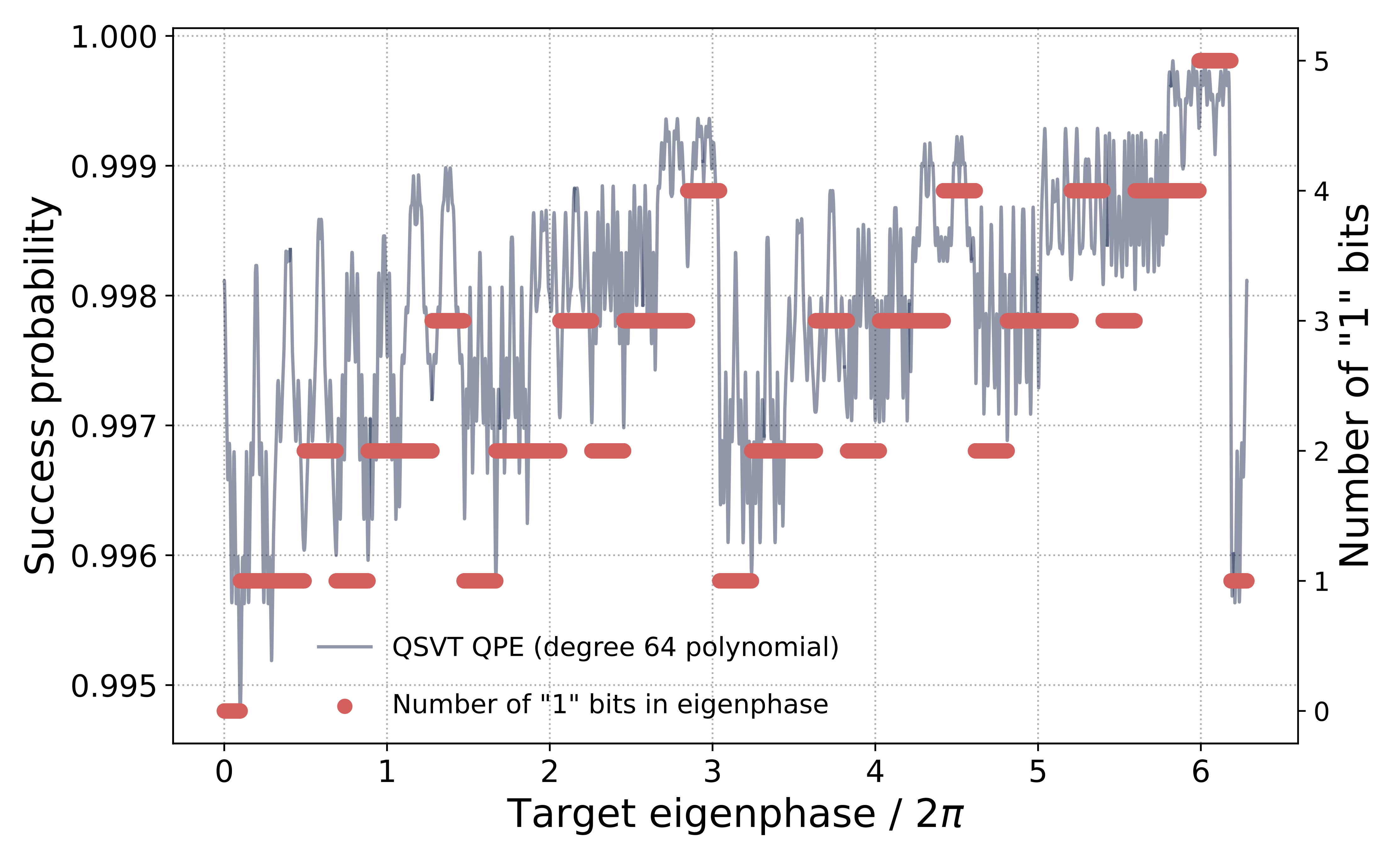}
\caption{Plot of QSVT QPE success probability as a function of target eigenphase, with the number of $1$ bits in the binary fixed point representation of that eigenphase shown as green points overlaid on the QSVT QPE data (red line). The asymmetry of the QSVT QPE success probability is well explained by the number of 1 bits, which is explained by the asymmetry in the target function as shown in Fig.~\ref{fig:QSVT_QPE_function_asymmetry}.}
\label{fig:QSVT_QPE_asymmetry}
\end{figure}

As a result of this asymmetry, the error in the QSVT QPE routine is dependent on the number of $1$ bits in the ideal eigenphase, which breaks the periodicity of the results. Fig.~\ref{fig:QSVT_QPE_asymmetry} shows this effect for statevector simulations of QSVT QPE with 5 phase qubits using a 64 degree polynomial approximation of the shifted sign function. Overlaid with the statevector simulation data (dark blue line) are the number of 1 bits in the closest 5-bit approximation to each ideal eigenphase (red points). The results of the statevector simulation match the number of 1 bits to a high degree, corroborating the explanation for the asymmetry of QSVT QPE.

\end{document}